\documentclass[twocolumn]{aastex62}

\usepackage{rotating}

\usepackage{graphicx}
\usepackage{epstopdf}
\usepackage{natbib}
\usepackage{xspace}
\usepackage{color}
\usepackage{amsmath}
\newcommand{\kms}     {\,km\,s$^{-1}$\xspace}

\newcommand{\mjy}     {\,mJy\,beam$^{-1}$\xspace}

\newcommand{\msun}    {\,M$_{\sun}$\xspace}

\newcommand{\cmd}     {\,cm$^{-2}$\xspace}
\newcommand{\cmt}     {\,cm$^{-3}$\xspace}

\newcommand{\target}     {G5.89--0.39\xspace}

\shorttitle{Linear Polarized Imaging of the UCHII Region \target}
\shortauthors{Fern\'andez L\'opez et al.}
\begin{document}

%\title{ALMA observations of the molecular environment surrounding DO\,Tauri}
\title{Magnetic Fields in Massive Star-Forming Regions (MagMaR) I. Linear Polarized Imaging of the UCHII Region \target}

\author[0000-0001-5811-0454]{M. Fern\'andez-L\'opez}
\affiliation{Instituto Argentino de Radioastronom\'\i a (CCT-La Plata, CONICET; CICPBA), C.C. No. 5, 1894, Villa Elisa, Buenos Aires, Argentina}
\email{manferna@gmail.com}

\author[0000-0002-7125-7685]{P. Sanhueza}
\affiliation{National Astronomical Observatory of Japan, National Institutes of Natural Sciences, 2-21-1 Osawa, Mitaka, Tokyo 181-8588, Japan}
\affiliation{Department of Astronomical Science, SOKENDAI (The Graduate University for Advanced Studies), 2-21-1 Osawa, Mitaka, Tokyo 181-8588, Japan}

\author[0000-0003-2343-7937]{L. A. Zapata}
\affiliation{Instituto de Radioastronom\'\i a y Astrof\'\i sica, Universidad Nacional Aut\'onoma de M\'exico, P.O. Box 3-72, 58090, Morelia, Michoac\'an, M\'exico}

\author[0000-0003-3017-4418]{I. Stephens}
\affiliation{Department of Earth, Environment and Physics, Worscester State University, Worcester, MA 01602, USA}
\affiliation{Center for Astrophysics $|$ Harvard \& Smithsonian, Cambridge, MA 02138, USA}

\author[0000-0002-8975-7573]{C. Hull}
\affiliation{National Astronomical Observatory of Japan, NAOJ Chile, Alonso de C\'ordova 3788, Office 61B, 7630422, Vitacura, Santiago, Chile}
\affiliation{Joint ALMA Observatory, Alonso de C\'ordova 3107, Vitacura, Santiago, Chile}

\author[0000-0003-2384-6589]{Q. Zhang}
\affiliation{Center for Astrophysics $|$ Harvard \& Smithsonian, Cambridge, MA 02138, USA}

\author{J. M. Girart}
\affiliation{Institut de Ci\`{e}ncies de l'Espai (ICE-CSIC), Campus UAB, Carrer de Can Magrans S/N, E-08193 Cerdanyola del Valles, Catalonia}
\affiliation{Institut d'Estudis Espacials de Catalunya, E-08030 Barcelona, Catalonia}

\author{P. M. Koch}
\affiliation{Academia Sinica Institute of Astronomy and Astrophysics, No.1, Sec. 4., Roosevelt Road, Taipei 10617, Taiwan}

\author{P. Cort\'es}
\affiliation{National Radio Astronomy Observatory, Charlottesville, VA 22903, USA}
\affiliation{Joint ALMA Observatory, Alonso de C\'ordova 3107, Vitacura, Santiago, Chile}

\author[0000-0001-9500-604X]{A. Silva}
\affiliation{National Astronomical Observatory of Japan, National Institutes of Natural Sciences, 2-21-1 Osawa, Mitaka, Tokyo 181-8588, Japan}

\author{K. Tatematsu}
\affiliation{Nobeyama Radio Observatory, National Astronomical Observatory of Japan, National Institutes of Natural Sciences, 462-2 Nobeyama, Minamimaki, Minamisaku, Nagano 384-1305, Japan}
\affiliation{Department of Astronomical Science, SOKENDAI (The Graduate University for Advanced Studies), 2-21-1 Osawa, Mitaka, Tokyo 181-8588, Japan}

\author[0000-0001-5431-2294]{F. Nakamura}
\affiliation{National Astronomical Observatory of Japan, National Institutes of Natural Sciences, 2-21-1 Osawa, Mitaka, Tokyo 181-8588, Japan}
\affiliation{Department of Astronomical Science, SOKENDAI (The Graduate University for Advanced Studies), 2-21-1 Osawa, Mitaka, Tokyo 181-8588, Japan}
\affiliation{The University of Tokyo, Hongo, Bunkyo, Tokyo 113-0033, Japan}

\author[0000-0003-0990-8990]{A. E. Guzm\'an}
\affiliation{National Astronomical Observatory of Japan, National Institutes of Natural Sciences, 2-21-1 Osawa, Mitaka, Tokyo 181-8588, Japan}

\author{Q. Nguyen Luong}
\affiliation{McMaster University, 1 James St N, Hamilton, ON, L8P 1A2, Canada }
\affiliation{Graduate School of Natural Sciences, Nagoya City University, Mizuho-ku, Nagoya, Aichi 467-8601, Japan}

\author{E. Guzm\'an Ccolque}
\affiliation{Instituto Argentino de Radioastronom\'\i a (CCT-La Plata, CONICET; CICPBA), C.C. No. 5, 1894, Villa Elisa, Buenos Aires, Argentina}

\author{Y.-W. Tang}
\affiliation{Academia Sinica Institute of Astronomy and Astrophysics, No.1, Sec. 4., Roosevelt Road, Taipei 10617, Taiwan}

\author[0000-0002-9774-1846]{H.-R. V. Chen}
\affiliation{Institute of Astronomy and Department of Physics, National Tsing Hua University, Hsinchu 30013, Taiwan}

%
%
%
%\author{Y.W. Tang}
%\affiliation{}
%
%\author{B. Commercon}
%\affiliation{}
%
%\author{P. Hennebelle}
%\affiliation{}
%
%\author{H-B Li}
%\affiliation{}
%
%\author{X. Lu}
%\affiliation{}
%
%\author{J. Jackson}
%\affiliation{}
%
%\author{Y. Contreras}
%\affiliation{}
%
%\author{T. Sakai}
%\affiliation{}
%
%\author{B. Wu}
%\affiliation{}
%
%
%\author{K. Saigo}
%\affiliation{}
%
%\author{T. Saito}
%\affiliation{}

\begin{abstract}
We report 1.2\,mm polarized continuum emission observations carried out with the Atacama Large Millimeter/submillimeter Array (ALMA) toward the high-mass star formation region \target. The observations show a prominent 0.2\,pc north-south filamentary structure. The UCHII in \target breaks the filament in two pieces. Its millimeter emission shows a dusty belt with a mass of 55-115\msun and 4,500\,au in radius, surrounding an inner part comprising mostly ionized gas with a dust emission only accounting about 30\% of the total millimeter emission. We also found a lattice of convex arches which may be produced by dragged dust and gas from the explosive dispersal event involving the O5 Feldt's star. The north-south filament has a mass between 300-600\msun and harbours a cluster of about 20 millimeter envelopes with a median size and mass of 1700\,au and 1.5\msun, respectively, some of which are already forming protostars.  

We interpret the polarized emission in the filament as mainly coming from magnetically aligned dust grains. The polarization fraction is $\sim4.4$\% in the filaments and $2.1$\% at the shell. The magnetic fields are along the North Filament and perpendicular to the South Filament. In the Central Shell, the magnetic fields are roughly radial in a ring surrounding the dusty belt between 4,500 and 7,500\,au, similar to the pattern recently found in the surroundings of Orion BN/KL. This may be an independent observational signpost of explosive dispersal outflows and should be further investigated in other regions.
%Finally, toward the center of the most massive dust millimeter envelope found (MM15), the magnetic fields change abruptly by $90\degr$, and the polarization fraction, which was decreasing progressively from the outskirts, has a sudden turn at its center.    
\end{abstract}


\section{Introduction} \label{s:intro}

Similar to their low-mass counterparts, high-mass stars form inside molecular clouds which span across parsecs in size, typically adopt filamentary shapes and can be seen as Infrared Dark Clouds \citep[IRDCs,][]{1994Lis,1996Perault,1998Carey,2000Carey,2005Hill,2006Rathborne,2006Simon,2019Sanhueza}. These filaments possibly originate by turbulence in the Interstellar Medium (ISM) imprinted at large scales by supernova explosions or high-speed massive star winds \citep[e.g.,][]{2015Inutsuka}, among other phenomena \citep[e.g., spiral density wave shocks, Parker and gravitational instabilities][]{1990Elmegreen}. Overdensities created within these structures fragment and collapse from dense cores down to 1,000\,au dusty envelopes scales, which harbor the formation of freshly new stellar systems \citep[e.g.,][]{2009Zhang,2010Bontemps,2015Zhang,2020Palau}, most of which are multiple systems \citep[e.g.,][]{2013Duchene,2017Moe}. It is well known that these young systems eject collimated winds or jets at velocities of 100-1000\kms, regulating the growth in the disk/envelope's angular momentum and allowing the accretion of gas and debris onto the protostars through rotating  circumstellar disks \citep[e.g.,][]{2007Cesaroni,2016Bally}. Some systems form objects several times more massive than the Sun, and launch very powerful jets transferring huge amounts of kinetic energy into the environment via dragging and shocking the molecular material of the originally quiescent filamentary cloud \citep[e.g.,][]{2002Beuther,2005Zhang,2008Qiu,2009LopezSepulcre,2013FernandezLopez,2019Zapata_iras16547}. Occasionally, gravitational interactions between the objects in these stellar systems can become chaotic. This occurs in systems with three or more objects orbiting each other. Some of these systems may end up with stars in very close orbits, which could contain colliding paths. In such occasions, it is plausible that the system is disintegrated, the individual disks tear into pieces, and sometimes, even new objects could be formed from a violent merger \citep{2005Bally,2005Rodriguez,2005Bonnell}.

%Al igual que las estrellas de baja masa, las estrellas de alta masa se forman usualmente dentro de nubes moleculares gigantes, con tama;os de parsecs, que poseen t[ipicamente formas filamentarias. Es posible que estos filamentos esten originados por la turbulencia aimprimida n el medio interestelar a gran escala por explosiones de superrnova o vientos de estrellas masivas, entre otros fenomenos. Las sobredensidades que se originan dentro de estas estructuras producen la fragmentaci[on y el colapso de nucleos que derivan en la formacion de nuevos sistemas estelares, que en su mayor parte son sistemas multiples. Es bien sabido que estos sistemas estelares jovenes eyectan vientos colimados o jets a grandes velocidades, lo que regula el aumento del momento angular, permitiendo la acrecion de material por parte de las protoestrellas a traves de sus discos circumestelares. 
%Algunos de estos sistemas forman objetos varias veces mas masivos que nuestro sol, y lanzan jets muy poderosos introduciendo grandes cantidades de energia cinetica en el medio circundante, arrastrando y chocando el material molecular que originalmente formaba parte de la nube filamentaria. En ocasiones, las interacciones gravitatorias de los objetos de estos sistemas pueden volverse caoticas en sistemas de mas de dos elementos. Algunos de estos sistemas pueden terminar con estrellas en orbitas muy cercanas o incluso colisiones que producen la desintegracion del sistema y/o la formacion de nuevos objetos en un proceso de fusion violenta.

The role played by the magnetic fields in each of the processes mentioned above is still not well understood because of the difficulty in their detection and measurement \citep[but see][for current ideas on this matter]{2019Hull}. The morphology of the field lines is also a difficult problem to solve since most of the times, only a 2D projection in the plane of the sky can be inferred. Spinning grains preferentially align with their elongated dimension perpendicular to the field lines. Hence the thermal millimeter and submillimeter emission from the grains is linearly polarized in a direction orthogonal to the magnetic field \citep[see e.g.,][]{1994Lazarian,2007Cho,2007Lazarian,2015Andersson}. Specifically, the magnetic field inside ISM filaments has been studied at parsec scales using observations with the Planck satellite \citep[e.g.,][]{2016PlanckCollaboration}, revealing that in the densest filaments (like those forming the most massive stars in the Galaxy) the main direction of the magnetic field is generally perpendicular to their main direction \citep[see also,][]{2017Pattle,2019Soam}. Within the parsec-scale massive clumps where protoclusters are expect to form, spatially resolved observations revealed magnetic field configurations ranging from a hourglass shaped morphology (e.g., G34.41+0.31: \citealt{2009Girart,2019Beltran}, and G240.31+0.07 \citealt{2014Qiu}) to a more random distribution \citep[e.g., DR21(OH)][]{2013Girart}. However, a systematic SMA polarimetric study of 14 high-mass star forming clumps found, on a statistical basis, that the magnetic field in dense cores tends to be aligned, and perpendicular to the field of their parental clump \citep{2014Zhang}. 

%El papel que juegan los campos magneticos en cada uno de estos procesos aun no esta bien entendido, debido a la dificultad de medirlos y de detectar su morfologia. Una de las formas que tenemos para investigarlos es a traves de experimentos de luz polarizada, ya que los spinning granos de polvo del medio interestelar, se alinean de forma parcial generalmente con sus lados elongados de forma perpendicular al las lineas de campo. La emision milimeterica y submilimetrica de dichos granos queda de este modo linealmente polarizada en direccion ortogonal al campo magnetico.  En particular, el campo magnetico en los filamentos del medio interestelar se ha estudiado a gran escala con observaciones del sat[elite Planck (e.g., Planck Collaboration et al. 2016a,b,c;Soler et al. 2017; Fissel et al. 2019), descubriendose que en los filamentos mas densos (aquellos en los que se forman las estrellas masivas) el campo magnetico esta preferentemente orientado de forma perpendicular a los mismos (Pattle et al.2017; Ward-Thompson et al. 2017; Liu et al. 2018; Soam et al. 2019). 

The Ultra-Compact HII region (UC\,HII) \target, also known as W28\,A2, is a region that has probably formed (or it is forming) a new generation of massive stars \citep[e.g.,][and references therein]{2019Zapata}. Maser emission from several molecules have been spotted in the region \citep[e.g., H$_2$O, CH$_3$OH, OH, NH$_3$,][]{1996Hofner,2007Stark,2004Kurtz,2008Hunter}. A huge amount of dense molecular material has been found surrounding the shell-like UCHII region, displaying hints of expansion motions \citep{1991Gomez}. Studying the ionized gas of the UCHII, \cite{1998Acord} derived that the HII region is expanding at a velocity of 35\kms and it has a dynamical age of 600$^{+250}_{125}$\,years. The gas in the UCHII region could be ionized by a young star with the luminosity of a O5V ZAMS star, identified by high-angular resolution near-infrared observations, and lying at the northeast edge of the shell \citep{2003Feldt}. This star may have been at the center of the UCHII region 1000\,years ago if moving at about 10\kms. \cite{2006Puga} inferred the presence of a possible second young star in the 
northwest edge of the shell through the discovery of a possible Br$\gamma$ ionized jet, but there is no direct evidence on the existence of such object yet. Several extremely powerful outflows in different directions were reported in \target \citep{1988Harvey,1990Zijlstra,2004Sollins,2008Hunter}. The kinetic energy of these ejections amounts $10^{46}-10^{48}$\,ergs, placing them among the most powerful outflows in the Galaxy. Very recently, tens of CO outflowing filaments have been found expanding from the center of the UCHII region confirming the existence of a new explosive outflow \citep{2019Zapata,2020Zapata}. The discovery may imply that this kind of events are much frequent than previously thought \citep{2020Zapata}. In addition, possibly related with the event of an explosive outflow, \target spatially matches with several high-energy sources from X-rays to TeV $\gamma$-rays, such as HESS\,J1800-240B \citep[e.g.,][]{2015Gusdorf,2016Hampton}. The TeV sources have been explained as the interaction between a population of cosmic-rays accelerated in the supernova remnant W28 (located $\sim1\degr$ north of \target) and the molecular envelope of the UCHII region \citep{1979Montmerle,2017Montmerle}. However, shocks produced by the explosive event may be strong enough to accelerate particles up to the relativistic regime. Continuum millimeter and submillimeter observations carried out previously with the Submillimeter Array detected the dust associated with the edges of the UCHII region and the two north-south elongations \citep{2004Sollins}. These structures were later resolved into a dusty ring fitting the UCHII region, and the emission from three additional sources to the north, east and south \citep{2008Hunter}. Moreover, submillimeter polarized emission was detected toward a ridge of dust extending north-south across the western edge of the shell \citep{2009Tang}. These observations served to set an upper limit to the magnetic field strength on the plane of the sky ($<2-3$\,mG) and to conclude that the field lines morphology are shaped by actors such as the radiation, the pressure and the outflows of the region.
 
In this work we adopt a distance to \target of 2.99$^{+0.19}_{-0.17}$\,kpc estimated by \cite{2014Sato} after measuring the proper motions and infer the parallaxes of a group of 22\,GHz H$_2$O masers. However, a smaller distance of 1.28\,kpc was obtained using the same technique \citep{2011Motogi}, and other works in the literature show a controversy with measurements between 1.3 and 7.0\,kpc. In addition, the Gaia\,DR2 data inside a projected 3\,arcminutes ring centered at \target, show very large G-band extinctions ($>1$\,magnitudes at less than 1000\,pc). This is possibly due to the proximity of this target with the direction of the Galactic Center. Inspecting the Gaia\,DR2 data, it is not clear if there is one or more clouds overlapped toward this line of sight. Actually, an analysis of the HI absorption line indicates a distance between 3 and 7\,kpc \citep{1990Zijlstra}, and the stellar counting and extinction studies in red giants of \cite{2012Foster} both obtain distances between 3.5 and 4.5\,kpc, closer to the distance found by \cite{2014Sato}.

The present observations comprise deep 1.2\,mm ALMA polarization observations toward the \target region. This target was observed as part of the Magnetic Fields in Massive Star-Forming Regions (MagMaR) survey that in total contains 30 sources. Details on the survey and source selection will be given in Sanhueza et al. (2021, in prep.). In Section \ref{sec:obs}, we describe the observations, along with 2\,cm VLA archival observations. Section \ref{s:results} shows the results, split into the raw continuum and the linear polarized emission at millimeter wavelengths.  We discuss the results in Section \ref{s:discussion} and summarize the main ideas in Section \ref{s:conclusions}.

\section{Observations} \label{sec:obs}
\subsection{252\,GHz ALMA data}
Observations were conducted on 2018 September 25 using ALMA Band~6 (1.2\,mm) under the \textit{MagMar} project (code 2017.1.00101.S; PI: P. Sanhueza). A total of 47 antennas were used with baselines ranging between 15 and 1400\,m. The PWV during the observations ranged between 1.1 and 1.4\,mm and the phase rms was below $0.31\degr$. 
The spectral setup contains five spectral windows: three of them (centered at 243.472, 245.472 and 257.472\,GHz) are intended for continuum emission with a spectral resolution of 1.953\,MHz ($\sim$2.4\kms) and total bandwidth of 1.875\,GHz each, while the other two are focused on resolving the H$^{13}$CO$^{+}$ (3-2) and HN$^{13}$C (3-2) spectral lines with a spectral resolution of 488.281 kHz (0.56\kms) and bandwidth of 234.38\,MHz. The continuum coverage was centered around 252\,GHz with a total bandwidth of 5.8\,GHz. 

%ALMA delivered the calibrated data that was manually flagged (an atmospheric absorption on one spectral window and some outlier amplitudes were removed). 
J1832-2039 was used as the phase calibrator, while J1924-2914 was adopted as the bandpass, the absolute flux scale and the polarization calibrator. About a 10\% of absolute uncertainty in the flux calibration is expected for ALMA Band\,6 observations\footnote{ALMA Cycle 8 Technical Handbook, Chapter 10.2.6.}. Regarding the polarization uncertainty, the measured D-terms due to instrumental cross-talking between the two orthogonal polarized signals received (instrumental polarization) are less than 5\%. 

The observations toward \target were intertwined with observations of other six scientific targets, which will be reported in future papers. The total time on source was 12 minutes. After the standard calibration, we selected only the line-free channels to make maps of the continuum emission. Continuum-only channels account for 72\% of the total bandwidth available. We run three iterations of phase-only self-calibration which improved the signal-to-noise ratio in the Stokes I image from 44 to 111. The images were Fourier transformed, deconvolved and restored using the CASA tclean task. The final Stokes I image has a synthesized beam of $0\farcs29\times0\farcs22$, PA=84.1$\degr$ (870\,au$\times$660\,au at the assumed distance) and an overall rms noise level of 1.1\mjy. For the Stokes Q, U and V images the rms noise level is 0.04\mjy. We also build images of the debiased linear polarized intensity (m$_l=\sqrt{Q^2+U^2}$), the linear polarization fraction (p$_l=m_l/I$, average uncertainty of 0.6\%), the electric vector position angle (EVPA$=0.5\arctan{(U/Q)}$, average uncertainty of $6\degr$).
 % and the circular polarization fraction (p$_c=V/I$, average statistical uncertainty of 0.2\%).      

The ALMA polarimetric capabilities for the Cycle\,6 observations included full linear and circular operations for single pointing, with a field of view limited to the inner 1/3 and 1/10 of the primary beam, respectively (about $7\farcs5$ and $2\farcs3$ out of $22\farcs9$, at the observing frequency).  

\subsubsection{ALMA Circular Polarized Continuum Emission}\label{circular}
At first glance, there appears to be a 5-10$\sigma$ Stokes\,V detection in the ALMA data. The Stokes\,V spatial distribution is not coincident with the Stokes\,I emission. However, ALMA can sometimes create artificial Stokes\,V emission due to its beam squint pattern. To determine if this is a real signal, we made a simulated map of the ALMA beam-squint pattern. We rotated it at the different parallactic angles covered by the observations, and multiplied each resulting map with the Stokes\,I spatial distribution. We then Fourier transformed each of these maps and concatenate all of them into a single file of visibilities. This file was then treated as the real visibilities observed by ALMA. The final cleaned image reproduced quite well the spatial distribution of the original Stokes\,V obtained from the ALMA observations, indicating that the 5-10$\sigma$ detection is most probably spurious.

\subsection{Ancillary 8.4\,GHz VLA data}\label{s:vladata}
%Luis Felipe's reduction
%beam: 0.56x0.39, -88deg
%rms: 0.082 mJy
%Observing freq = 8.4399GHz
The observations were made with the Very Large Array of NRAO\footnote{The National 
Radio Astronomy Observatory is a facility of the National Science Foundation operated
under cooperative agreement by Associated Universities, Inc.} centered at a rest frequency of 8.4\,GHz (3.5\,cm) during
February 1989. At that time the array was in its AB configuration. The observations used the 27 antennas of the array.  
The absolute amplitude calibrator was 3C286, while the gain-phase calibrators were J1730$-$130 and J1741$-$038. 
The digital correlator of the VLA was configured in 2 spectral windows of 50 MHz. The data were analyzed in the standard manner using the AIPS package of NRAO. The resulting image {\it rms}-noise and angular resolution are 82 $\mu$Jy, and 0.56$''$ $\times$0.39$''$ with a position angle of $-$87.5$^\circ$, respectively. The field of view is $5\farcm4$. 

\begin{figure*}
\centering
\includegraphics[angle=0, width=\linewidth]{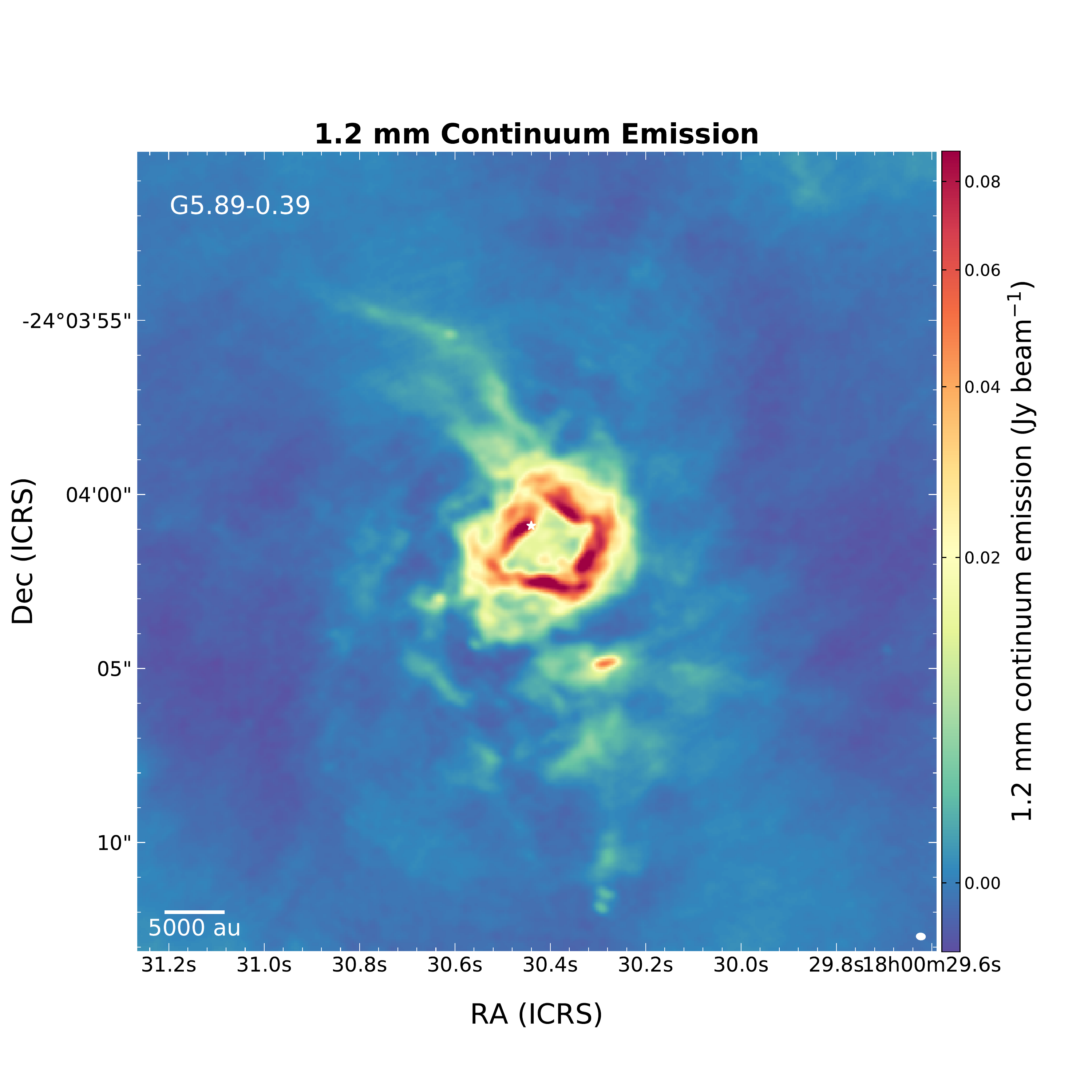}
\caption{1.2\,mm Stokes I continuum emission toward \target. The white symbol marks the position of the Feldt's star. The synthesized beam, in the bottom left corner, is $0\farcs29\times0\farcs22$, PA=84.1$\degr$. The rms noise level of the image is estimated in 1.1\mjy.}
\label{f:cont} 
\end{figure*}

\begin{figure*}
\centering
\includegraphics[angle=0, width=0.8\linewidth]{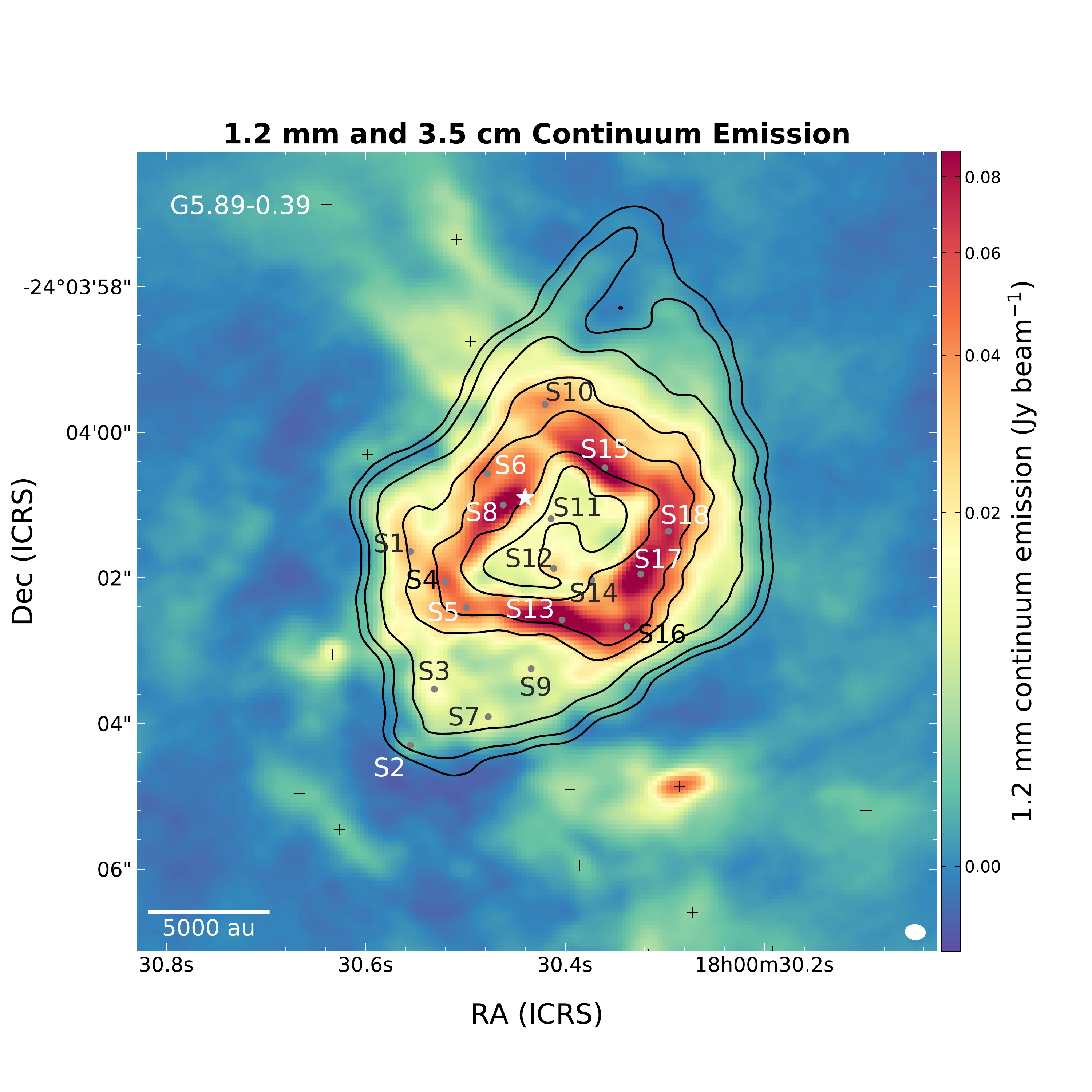}
\caption{Zoom in to the Central Shell structure. The 3.5\,cm VLA continuum emission in contours (at 2, 6, 24, 72 and 100 times the rms noise level of 82\,$\mu$m) is overlapped with the 1.2\,mm ALMA continuum emission (colour-scale). The free-free ionized emission contribution at 1.2\,mm toward the Central Shell is evidenced by the quite good overall spatial agreement of the emission at both wavelengths. That includes the inner shell's decrease of emission.
}
\label{f:free-free}
\end{figure*}

\begin{figure}
\centering
\includegraphics[angle=0, width=1.1\columnwidth]{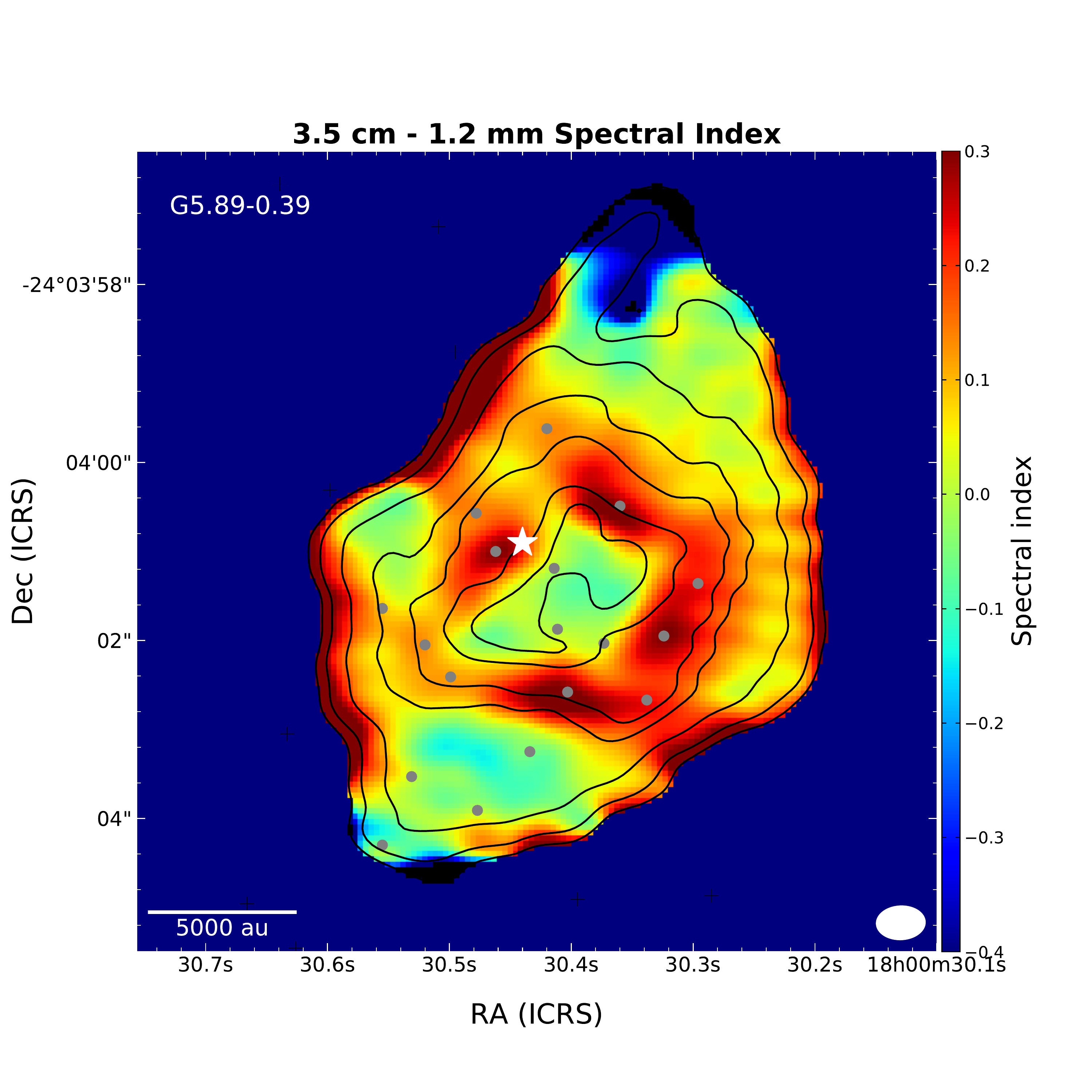}
\caption{Spectral index derived from the 3.5\,cm VLA and the 1.2\,mm ALMA continuum emission images. The VLA continuum emission is shown in contours at 2, 6, 24, 72 and 100 times the rms noise level. The spectral index $\alpha$ is estimated using $S_{\nu}\propto\nu^{\alpha}$.
}
\label{f:spix}
\end{figure}

\section{Results}\label{s:results}
In this section we show the main results derived from the continuum ALMA observations toward \target. First, we present a description of the main features of the Stokes\,I continuum emission at 1.2\,mm in this region, and second, we introduce the main linear polarization results.

%%%%%%%% Stokes I
\subsection{1.2\,mm Stokes\,I Continuum Emission}
\label{s:cont}
Figures \ref{f:cont}, \ref{f:free-free}, \ref{f:spix}, \ref{f:overview} and \ref{f:zoomin} show the overview of the millimeter emission toward \target, the comparison between the spatial distributions of the millimeter and centimeter emissions, the spectral index between both wavelengths, the location of the identified millimeter sources, and the detailed structure of the central part. We first address the large scale structures in the images (see also Table \ref{t:masses}), then we present the emission from the compact sources (Tables \ref{t:cores}, \ref{t:condensations}), and finally we identify the complex structures found at the center of the field of view. 
%We describe the emission of the main structures, measure fluxes, and calculate masses in the following.  

\subsubsection{Large Scale Structure}
\label{s:large_structure}
Overall, the emission is in agreement with previous lower angular resolution millimeter observations \citep{2008Hunter,2009Tang}, but our observations reveal a lot of more details. The most striking feature of the large scale structure is the strong emission from the shell-shaped structure toward the central HII region (Fig. \ref{f:cont}). The Central Shell matches the general resemblance of previous multiwavelength observations and the archival VLA 3.5\,cm emission, thus confirming the good agreement between the spatial distributions of both centimeter and millimeter observations \citep{2008Hunter}. In fact, if we zoom-in on the central structure (Fig. \ref{f:free-free}) and overlay the VLA\,3.5\,cm contours, we see that part of the millimeter emission could be produced by free-free. 
The shell emission comprises various elongated condensations S1--S18 (identified using a dendrogram analysis, see section \ref{s:cores} below, and marked as grey dots in Figs. \ref{f:free-free} and \ref{f:overview}, see also Table \ref{t:condensations}). In particular, the peak emission at 1.2\,mm in the whole region is located at S8 (123\mjy or 37\,K in brightness temperature, taking into account the beam size of the observations). As a reference point, let us mention that the O5\,V star \citep[hereafter the Feldt's star,][]{2003Feldt} lies to the northeast of the structure (marked by a white star), close to the position of the S8 condensation. Up to the location of this condensation, the shell-like structure has a radius of about $1\farcs5$ (about 4500\,au at the assumed distance) and shows an irregular shape, not smoothly circular, but more squared. The irregularity in the shape of the shell was noticed before by \cite{2006Puga}. 

As seen in Figure \ref{f:cont}, a prominent filament of dust extends due northeast from the Central Shell (hereafter the North Filament). It bifurcates and the northernmost tip forms a curved arch to the east. This filament has an approximate size of about $7\farcs0$ (0.10\,pc) and covers a range of position angles between $25\degr$ and $40\degr$ (measured from the center of the Central Shell). The filament width, close to the Central Shell is about $2\farcs5$, while farther north the west subfilament becomes narrower than $0\farcs7$ (about 2100\,au).  

Another filamentary structure extends also due south (the South Filament). It may well be the continuation of the North Filament if there were not an interruption by the Central Shell. The size of the South Filament is $8\farcs2$ (0.12\,pc) with a position angle of $8\degr$ (measured from the center of the Central Shell) and, as in the North Filament, the ALMA image shows several millimeter sources along its extent.

We estimate the total gas and dust masses of the large-scale structures that we have identified so far: the North Filament, the South Filament and the Central Shell. We assume optically thin isothermal dust emission and ignore the scattering opacity. Therefore, all the masses presented should be treated as well-informed lower limits. We use the relationship: 
$$M= \frac{d^2 F_{\nu}}{B_{\nu}(T_d)\kappa_{\nu}},$$
where $F_{\nu}$, d, $\kappa_{\nu}$ and $B_{\nu}(T_d)$, are the total observed flux density, the distance, the grain opacity (that depends on frequency as $\kappa_0\cdot(\nu/\nu_0)^{\beta}$), and the brightness of a black body at the dust temperature $T_d$, respectively. We choose a $\beta=2.0$, more appropriate for the dust of the Interstellar Medium \citep{1983Hildebrand}, resulting in $\kappa_{\nu}=1.072$\,cm$^2$~g$^{-1}$ at the observing frequency and appropriate for dust with thin ice mantles and denisites of $10^6$\cmt \citep[][]{1994Ossenkopf}. We assume a gas-to-dust ratio of 100 for the calculation. We also choose a cold dust temperature of 10-15\,K \citep{2014Lee} for the filaments and 75-150\,K for the Central Shell as discussed in \cite{2008Hunter}. The masses of the North and South Filaments are 125 and 145\msun when using T$_d=15$\,K (Table \ref{t:masses}). For a 10\,K temperature the derived masses are about a factor of two larger. 

\begin{deluxetable}{lccc}
\tablewidth{0pt}
\tablecolumns{4}
%tabletypesize{\scriptsize}
\tablecaption{Masses of the \target Large-Scale Structures}
\tablehead{
\colhead{Zone} & \colhead{Flux} & \colhead{T$_d$\tablenotemark{*}} & \colhead{Mass\tablenotemark{*}}\\
\colhead{} & \colhead{[mJy]} & \colhead{[K]} & \colhead{[\msun]} } 
\startdata
North Filament & 710 & 15/10 & 125/240 \\
Central Shell & 7488\tablenotemark{\dag} & 150/75 & 55/115 \\
South Filament & 820 & 15/10 & 145/275 \\
\enddata 
\tablecomments{The typical statistical uncertainties of the lower limit mass estimates due to the flux scale calibration are 10\%. However, the dominant uncertainty in these mass lower limit measurements is the temperature uncertainty.}
\tablenotetext{*}{Lower limit masses are derived assuming two different temperatures separated by the slash symbol.}
\tablenotetext{\dag}{Total flux measured at 252\,GHz in this work. We estimate that 2.959\,Jy of this flux may be due to free-free emission and subtracted it to estimate the mass (see section \ref{s:large_structure}).}
\label{t:masses}
\end{deluxetable}

\begin{figure}
\begin{center}
\includegraphics[width=\linewidth]{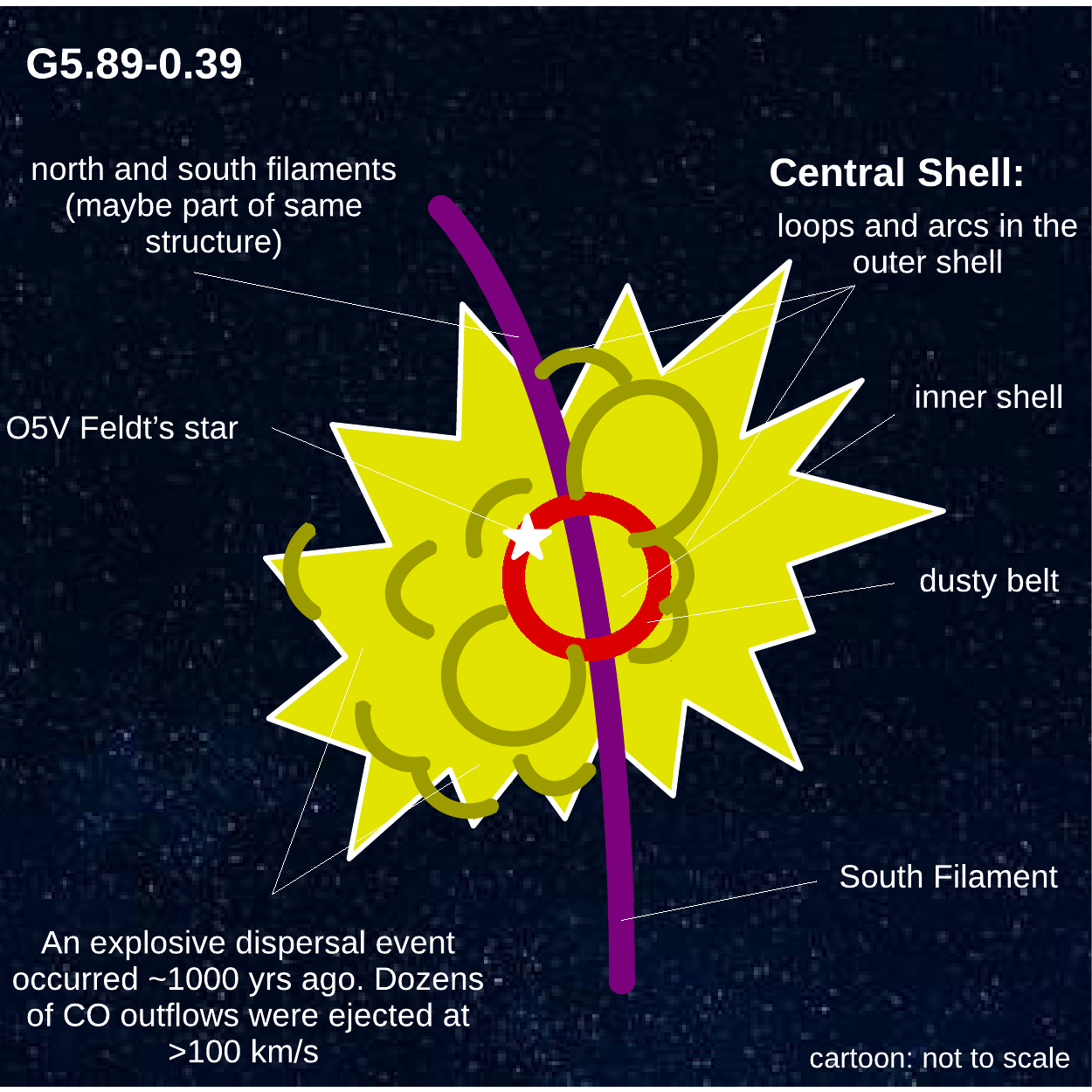}
\caption{Cartoon of the \target star-forming region showing the different structures found in the ALMA 1.2\,mm continuum emission, plus an idealized representation of the CO explosive dispersal outflow reported by \citep{2020Zapata}.}
\label{f:cartoon}
\end{center}
\end{figure}

Now, due to the contamination of free-free emission at millimeter wavelengths toward the Central Shell, we first need to subtract this contribution before calculating its total gas and dust mass. We follow the analogous estimate by \cite{2008Hunter} and adopt for the shell emission a uniform fiducial free-free radio spectral index of $\alpha=-0.154$, adequate for the optically thin emission in \target (here, $S_{\nu}\propto\nu^{\alpha}$). The millimeter flux emission measured from the ALMA image at 252\,GHz from the whole shell is 7.488\,Jy, and from the VLA image at 8.4\,GHz we extract a flux density of 4.992\,Jy. The latter centimeter flux density, extrapolated using the fiducial value of $\alpha$ to 252\,GHz, becomes 2.959\,Jy. This amounts for 40\% of the total flux measured at 252\,GHz. The thermal dust emission at 252\,GHz is therefore 4.529\,Jy, resulting in a mass between 55 and 115\msun, depending on the dust temperature chosen\footnote{The mass of the two filaments and the Central Shell together, roughly agrees with the 300\msun gas mass estimated by \cite{2009Tang} for the whole complex.}. Moreover, we analyze the emission from the Central Shell splitting it in four different zones (see Fig. \ref{f:zoomin}): we first consider (i) an inner shell, and, increasing in radius from the center, (ii) a belt of strong continuum emission, and (iii) an external zone containing both an ensemble of arches and bow-shaped features and, further away, (iv) more wispy and diffuse emission from larger loops and arcs (see Fig. \ref{f:cartoon}). 
We estimate the actual measured spectral index of the three innermost zones using the VLA and ALMA observations (Fig. \ref{f:spix}). We also calculate their fraction of free-free emission at 252\,GHz in the same manner as discussed above. 
In the inner shell, 77\% of the emission comes from a gas with a spectral index consistent with that of an optically thin free-free ionized gas ($\alpha=-0.08$, note however that measurements at three different wavelengths are needed to determine the opacity of the free-free emission) and we estimate it contains 1.4/2.9\msun of gas and dust mass. The dusty belt shows a rising spectral index ($\alpha=0.18$) suggesting 30\% of free-free contamination. It contains 52-108\msun of gas, the bulk of the mass of the Central Shell.  Beyond the belt, in the external zone showing centimeter emission, the northwest and southeast protuberances show larger free-free contribution ($\sim67$\% and $\alpha=-0.04$) than the east and west part, which show 40\% free-free contribution and a derived spectral index $\alpha=0.10$ (Fig. \ref{f:spix}). The gas mass content in this region amounts 1.6/4.1\msun. Therefore the free-free contamination is dominant in the emission from the inner zone and the northwest and southeast external protuberances. This emission is consistent with optically thin ionized gas. In the belt and the external east and west parts the emission is dominated by dust with a smaller free-free contamination.

\subsubsection{Identification of Millimeter Envelopes and Dust Condensations}
\label{s:cores}
\begin{figure*}
\centering
\includegraphics[angle=0, scale=0.3,width=0.8\linewidth]{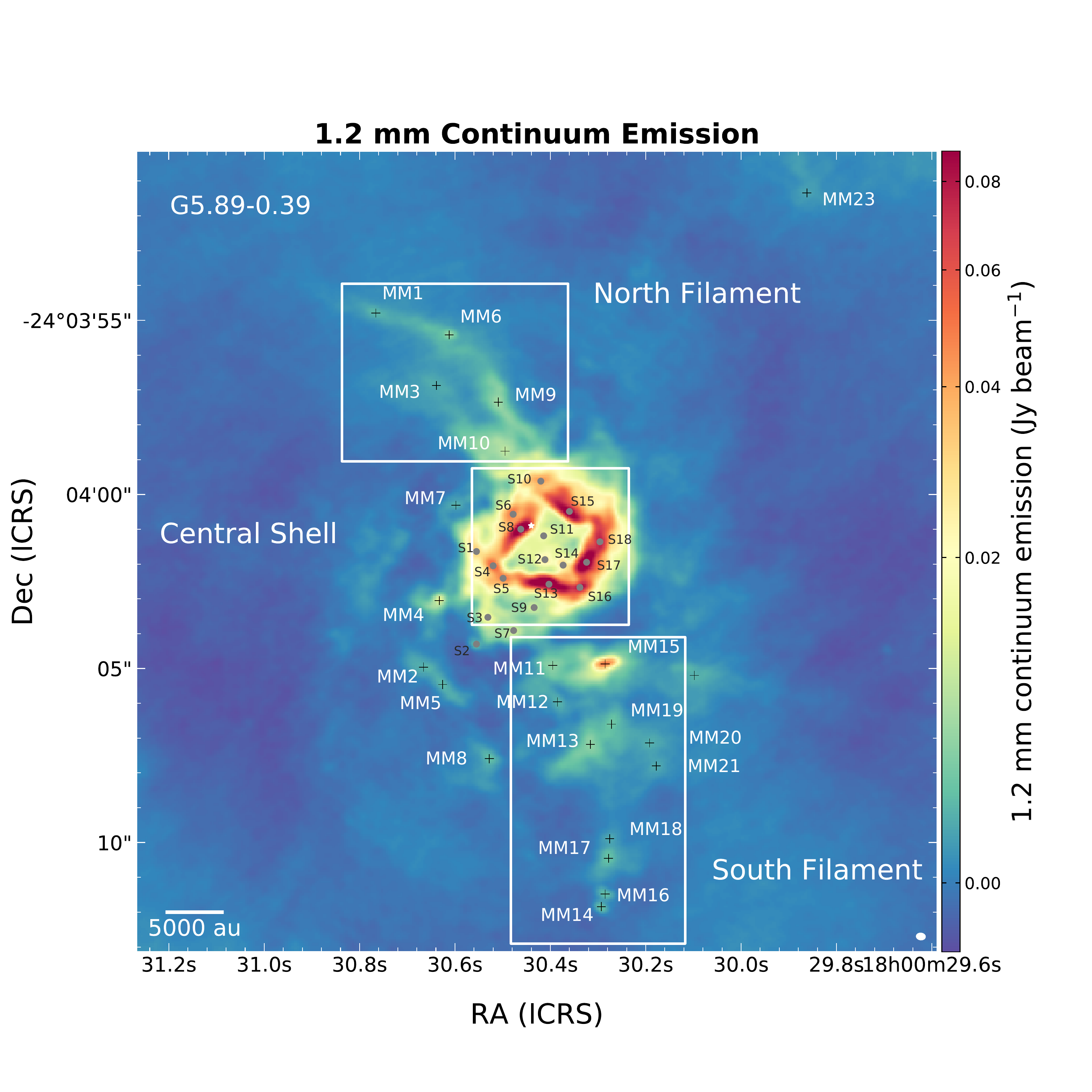}
\caption{1.2\,mm Stokes I continuum emission toward \target.  
The three main regions (North and South Filaments and Central Shell) are marked along with the main dusty condensations associated with the Central Shell (S1-S18) and the dusty envelopes identified via dendrograms (MM1-MM23).}
\label{f:overview} 
\end{figure*}

Using a dendrogram technique\footnote{The implementation of the algorithm used allows to set three parameters: (i) an intensity threshold (set to 1.8\mjy, which is 3.0 times the rms level of 0.6\mjy found locally in several parts of the image), (ii) a contrast factor to separate adjacent leaves (set to 0.5), and (iii) a minimum number of pixels for a leaf to be defined (set to the number of pixels inside one beam).} \citep{2008Rosolowsky} we have identified 22 millimeter envelopes\footnote{We identify these sources as dusty envelopes surrounding young stellar systems since their average size is $1750$\,au \citep[e.g.,][]{2009Zhang,2019Hull}, almost an order of magnitude smaller than the size of typical millimeter cores, but still much larger than circumstellar disks.}
toward the filamentary structure of \target, 18 dust condensations\footnote{For this work we define the millimeter envelopes as those sources undetected in the current VLA images. Likewise, we define the dust condensations as sources detected both at centimeter and millimeter wavelengths.} delineating the edge of the Central Shell or associated with it, and another envelope farther northwest of the Central Shell (MM23), probably not linked  with the main \target structure. In the rest of this work we refer to dusty envelopes or just envelopes to name the sources only detected at millimeter wavelengths, while we use dust condensations or condensations to name the sources with a mixture of millimeter and centimeter emission.
In Fig. \ref{f:overview} the millimeter envelopes are marked with black crosses and the condensations with grey dots.  The condensations show some degree of free-free contamination and are mainly detected toward the Central Shell emission. 
  
Five envelopes are located along the North Filament and 13 more along the South Filament 
(Table \ref{t:cores}). In the North Filament the envelopes are spread out along the filament separated by about $2\arcsec$ between them. In the South Filament, the envelopes are crowded around three locations also separated by about $2\arcsec$ between them: the surroundings of MM15, previously known as SMA-S \citep[north part of the South Filament,][]{2008Hunter}, the surroundings of MM13 (middle part of the filament), and the surroundings of MM17 (southern tip of the filament). In addition, we identify four envelopes in the outskirts of the Central Shell, that are possibly related with it (MM2, previously known as SMA-E, MM4, MM5 and MM7).

\begin{deluxetable*}{lcccccc}
\tablewidth{0pt}
\tablecolumns{7}
\tabletypesize{\scriptsize}
\tablecaption{Millimeter Envelopes in \target}
\tablehead{
\colhead{Source} & RA & DEC & Deconvolved Size & Peak Intensity & Flux Density & Mass   \\
\colhead{}  & \colhead{ICRS} & \colhead{ICRS} & \colhead{$\arcsec\times\arcsec$,$\degr$} & \colhead{mJy\,beam$^{-1}$} & \colhead{mJy} & \colhead{\msun}}
\startdata
MM1 & 18:00:30.766 & -24:03:54.79 & 0.86$\pm$0.04$\times$0.38$\pm$0.02, 70$\pm$2  & 3.3$\pm$0.1 & 20$\pm$1  & 1.3/3.0 \\
MM2\tablenotemark{*} & 18:00:30.666 & -24:04:04.96 & 1.20$\pm$0.07$\times$0.49$\pm$0.03, 65$\pm$2  & 3.0$\pm$0.2 & 31$\pm$2 & 1.9/4.6 \\
MM3 & 18:00:30.639 & -24:03:56.87 & 1.79$\pm$0.08$\times$0.70$\pm$0.03, 60$\pm$2  & 2.7$\pm$0.1 & 56$\pm$2 & 3.5/8.3 \\
MM4/SMA-E\tablenotemark{*} & 18:00:30.633 & -24:04:03.05 & 0.33$\pm$0.04$\times$0.22$\pm$0.04, 144$\pm$24  & 14$\pm$1 & 34$\pm$3 & 2.1/5.0 \\
MM5\tablenotemark{*} & 18:00:30.626 & -24:04:05.46 & 0.94$\pm$0.04$\times$0.25$\pm$0.02, 39$\pm$1  & 3.3$\pm$0.1 & 18$\pm$1 & 1.1/2.7 \\
MM6 & 18:00:30.612 & -24:03:55.42 & 0.43$\pm$0.03$\times$0.30$\pm$0.02, 80$\pm$8  & 6.5$\pm$0.5 & 19$\pm$1 & 1.2/2.8 \\
MM7\tablenotemark{*} & 18:00:30.598 & -24:04:00.31 & 0.46$\pm$0.03$\times$0.24$\pm$0.02, 2.9$\pm$0.1  & 8.2$\pm$0.4 & 68$\pm$4 & 4.3/10.0 \\
MM8 & 18:00:30.528 & -24:04:07.59 & 0.57$\pm$0.05$\times$0.34$\pm$0.04, 32$\pm$8  & 3.7$\pm$0.2 & 16$\pm$1 & 1.0/2.4 \\
MM9 & 18:00:30.509 & -24:03:57.35 & 0.90$\pm$0.02$\times$0.55$\pm$0.01, 20$\pm$2  & 7.4$\pm$0.4 & 67$\pm$1 & 4.2/9.9 \\
MM10/SMA-N & 18:00:30.495 & -24:03:58.76 & 1.1$\pm$0.1$\times$0.6$\pm$0.06, 40$\pm$5  & 10.0$\pm$0.3 & 66$\pm$2 & 4.1/9.7 \\
MM11 & 18:00:30.395 & -24:04:04.91 & 0.85$\pm$0.03$\times$0.81$\pm$0.03, 179$\pm$50  & 8.2$\pm$0.4 & 75$\pm$2 & 4.7/11.1 \\
MM12 & 18:00:30.385 & -24:04:05.96 & 0.75$\pm$0.05$\times$0.35$\pm$0.03, 38$\pm$4  & 3.0$\pm$0.2 & 16$\pm$1 & 1.0/2.4 \\
MM13 & 18:00:30.316 & -24:04:07.18 & 1.0$\pm$0.04$\times$0.66$\pm$0.03, 178$\pm$4  & 6.4$\pm$0.2 & 74$\pm$3 & 4.6/10.9 \\
MM14 & 18:00:30.293 & -24:04:11.84 & 0.23$\pm$0.08$\times$0.11$\pm$0.09, 24$\pm$33  & 4.1$\pm$0.2 & 6$\pm$1 & 0.4/0.9 \\
MM15/SMA-S & 18:00:30.285 & -24:04:04.87 & 0.60$\pm$0.03$\times$0.21$\pm$0.02, 104$\pm$2  & 47$\pm$3 & 163$\pm$7  & 10.2/24.1 \\
MM16 & 18:00:30.285 & -24:04:11.48 & 0.37$\pm$0.04$\times$0.16$\pm$0.03, 55$\pm$6  & 4.3$\pm$0.2 & 8$\pm$0.6 & 0.5/1.2 \\
MM17 & 18:00:30.278 & -24:04:10.45 & 0.71$\pm$0.03$\times$0.47$\pm$0.02, 161$\pm$5  & 4.2$\pm$0.1 & 24$\pm$1 & 1.5/3.5\\
MM18 & 18:00:30.275 & -24:04:09.89 & 0.44$\pm$0.01$\times$0.33$\pm$0.01, 125$\pm$5  & 2.4$\pm$0.1 & 8$\pm$1 & 0.5/1.2 \\
MM19 & 18:00:30.272 & -24:04:06.60 & 1.04$\pm$0.02$\times$0.56$\pm$0.01, 154$\pm$1  & 4.4$\pm$0.2 & 47$\pm$1 & 2.9/6.9 \\
MM20 & 18:00:30.192 & -24:04:07.14 & 0.79$\pm$0.02$\times$0.67$\pm$0.02, 110$\pm$9  & 2.8$\pm$0.1 & 26$\pm$1 & 1.6/3.8 \\
MM21 & 18:00:30.178 & -24:04:07.80 & 0.68$\pm$0.04$\times$0.53$\pm$0.04, 142$\pm$12  & 2.5$\pm$0.1 & 16$\pm$1 & 1.0/2.4 \\
MM22 & 18:00:30.098 & -24:04:05.20 & 1.8$\pm$0.1$\times$0.79$\pm$0.04, 77$\pm$3  & 3.0$\pm$0.2 & 68$\pm$4 & 4.3/10.0 \\
MM23/SMA-W & 18:00:29.862 & -24:03:51.34 & 0.63$\pm$0.02$\times$0.51$\pm$0.01, 85$\pm$6  & 2.1$\pm$0.1 & 13$\pm$1  & 0.8/1.9 \\
\enddata 
\tablecomments{Masses estimated using dust temperatures 40 and 20\,K are separated by a slash and presented in this order.}
\tablenotetext{*}{Millimeter envelope possibly associated with the Central Shell but undetected in the VLA observations.}
\label{t:cores}
\end{deluxetable*}
 
To estimate lower limits for the total gas and dust masses of the envelopes, we again assume that their emission is optically thin and isothermal, and we ignore scattering. We adopt a dust temperature between 20 and 40\,K, a more appropriate $\beta=1.5$ for dust in denser regions \citep[see e.g.,][]{2014Lee}, resulting in a value of $\kappa_{\nu} = 1.026$\,cm$^2$~g$^{-1}$ at 1.2\,mm \citep{1994Ossenkopf}. Table \ref{t:cores} contains the estimations of the masses of all the millimeter envelopes in the region based upon integrated flux densities from 2D-Gaussian fits to their millimeter emission. Using T$_{d}=40\,K$, the masses range between 0.4 and 10.2\msun with an average value of 2.4\msun and a median of 1.5\msun (masses are about a factor of two larger when considering the lower temperature of 20\,K).
The most massive envelope (MM15) is located $3\farcs5$ south of the center of the HII region and other 6 envelopes show intermediate masses of 3--5\msun (MM3, MM7, MM9, MM10, MM13 and MM22).

Several condensations, observed at millimeter and centimeter wavelengths, demarcate the strongest emission from the Central Shell. We have identified ten of them along the dusty belt, another three of them (S11, S12 and S14) within the shell, one (S1) southwest of it, within a small partially ionized gas arch, and four more (S2, S3, S7 and S9) associated with a prominent loop.  
The mass estimates of these condensations require a more careful treatment, since we first have to calculate the fraction of free-free emission at millimeter wavelengths for each of them. In this case we measure the flux density at both 8.4 and 252\,GHz, integrating within boxes defined manually by following the millimeter spatial distribution of every source. Table \ref{t:condensations} includes these fluxes along with calculated $\alpha$ spectral indices and the fraction of free-free contamination in the ALMA 1.2\,mm, derived by extrapolating the centimeter emission (and using a spectral index of $-0.154$) for every source \citep{2008Hunter,2009Tang}. Only seven sources have a free-free contamination fraction over 0.50. These are the sources that lie both inside (S11, S12 and S14) and outside of the dusty belt of the shell (Figs. \ref{f:free-free} and \ref{f:overview}), forming the southwestern prominent loop of ionized gas mentioned before (S2, S3, S7 and S9). In addition, all sources but two have a flat or slightly positive spectral index consistent with the model used by \cite{2008Hunter} of an HII region with a frequency turnover below 30\,GHz, plus a thermal dust emission component. S9 and S14 show, however, a negative spectral index indicating that their millimeter emission mainly stems from pure ionized gas.     

Using a dust temperature of 150\,K, the derived masses of the condensations S1--S18 range between 0.03 and 8.6\msun (Table \ref{t:condensations}), with a mean and median values of 2.0 and 0.9\msun, respectively. We obtain a factor of two larger masses when using a lower temperature of 75\,K.

\begin{deluxetable*}{lccccccc}
\tablewidth{0pt}
\tablecolumns{7}
\tabletypesize{\scriptsize}
\tablecaption{Condensations associated with the Central Shell}
\tablehead{
\colhead{Source} & RA & DEC & S$_{\textrm{8.4GHz}}$ & S$_{\textrm{252GHz}}$ & $\alpha$ & Fraction of & Mass   \\
\colhead{}  & \colhead{ICRS} & \colhead{ICRS} & \colhead{mJy} & \colhead{mJy} & & free-free & \colhead{\msun}}
\startdata
S1* & 18:00:30.555 & -24:04:01.64 & 41 & 64  & 0.1 & 0.38 & 0.6/1.2 \\
S2\tablenotemark{*} & 18:00:30.555 & -24:04:04.30 & 4 & 4 & 0.0 & 0.59 & 0.03/0.05 \\
S3\tablenotemark{*} & 18:00:30.531 & -24:04:03.53 & 56 & 65  & 0.0 & 0.51 & 0.5/1.0 \\
S4 & 18:00:30.520 & -24:04:02.05 & 98 & 162  & 0.1 & 0.36 &  1.6/3.2 \\
S5 & 18:00:30.499 & -24:04:02.41 & 42 & 72  & 0.2 & 0.35 & 0.7/1.5 \\
S6 & 18:00:30.478 & -24:04:00.57 & 61 & 106  & 0.2 & 0.34 & 1.0/2.2 \\
S7\tablenotemark{*} & 18:00:30.477 & -24:04:03.91 & 28 & 33  & 0.0 & 0.50 & 0.2/0.5 \\
S8 & 18:00:30.462 & -24:04:01.00 & 147 & 363  & 0.3 & 0.24 & 3.1/8.6  \\
S9\tablenotemark{*} & 18:00:30.434 & -24:04:03.25 & 33 & 27  & -0.1 & 0.72 & 0.1/0.2 \\
S10 & 18:00:30.420 & -24:03:59.62 & 69  & 114  & 0.1 & 0.36 & 1.1/2.3 \\
S11\tablenotemark{\dag} & 18:00:30.414 & -24:04:01.19 & 48 & 44 & -0.0 & 0.65 & 0.2/0.5 \\
S12\tablenotemark{\dag} & 18:00:30.411 & -24:04:01.87  & 45 & 52 & 0.0 & 0.51 & 0.4/0.8 \\
S13 & 18:00:30.403 & -24:04:02.58 & 255 & 670 & 0.3 & 0.23 & 7.7/16.1 \\
S14\tablenotemark{\dag} & 18:00:30.373 & -24:04:02.03 & 33 & 17 & -0.2 & 1.15 & \nodata\tablenotemark{\ddag}  \\
S15/SMA1 & 18:00:30.360 & -24:04:00.49 & 206 & 476 & 0.2 & 0.26 & 5.2/11.0 \\
S16 & 18:00:30.338 & -24:04:02.67 & 73 & 171 & 0.2 & 0.25 & 1.9/4.0 \\
S17/SMA2 & 18:00:30.324 & -24:04:01.95 & 197 & 691 & 0.4 & 0.17 & 8.6/17.9 \\
S18 & 18:00:30.296 & -24:04:01.36 & 88 & 221 & 0.3 & 0.24 & 2.5/5.3 \\
\enddata 
\tablecomments{Masses of the S-condensations in the shell are estimated after subtraction of a free-free contribution extrapolated from the 3.5\,cm emission with a spectral index of $\alpha=-0.154$. They are estimated using dust temperatures 150 and 75\,K, separated by a slash in the last column and presented in this order.}
\tablenotetext{*}{Compact condensation external to the Central Shell.}
\tablenotetext{\dag}{Condensation which lies, in projection, inside the Central Shell.}
\tablenotetext{\ddag}{\,Only ionized gas is found in this object at 1.2\,mm.}
\label{t:condensations}
\end{deluxetable*} 

\subsubsection{Arches and Loops Toward the Central Shell}
\label{s:arches}
\begin{figure*}
\begin{center}
\includegraphics[width=0.8\linewidth]{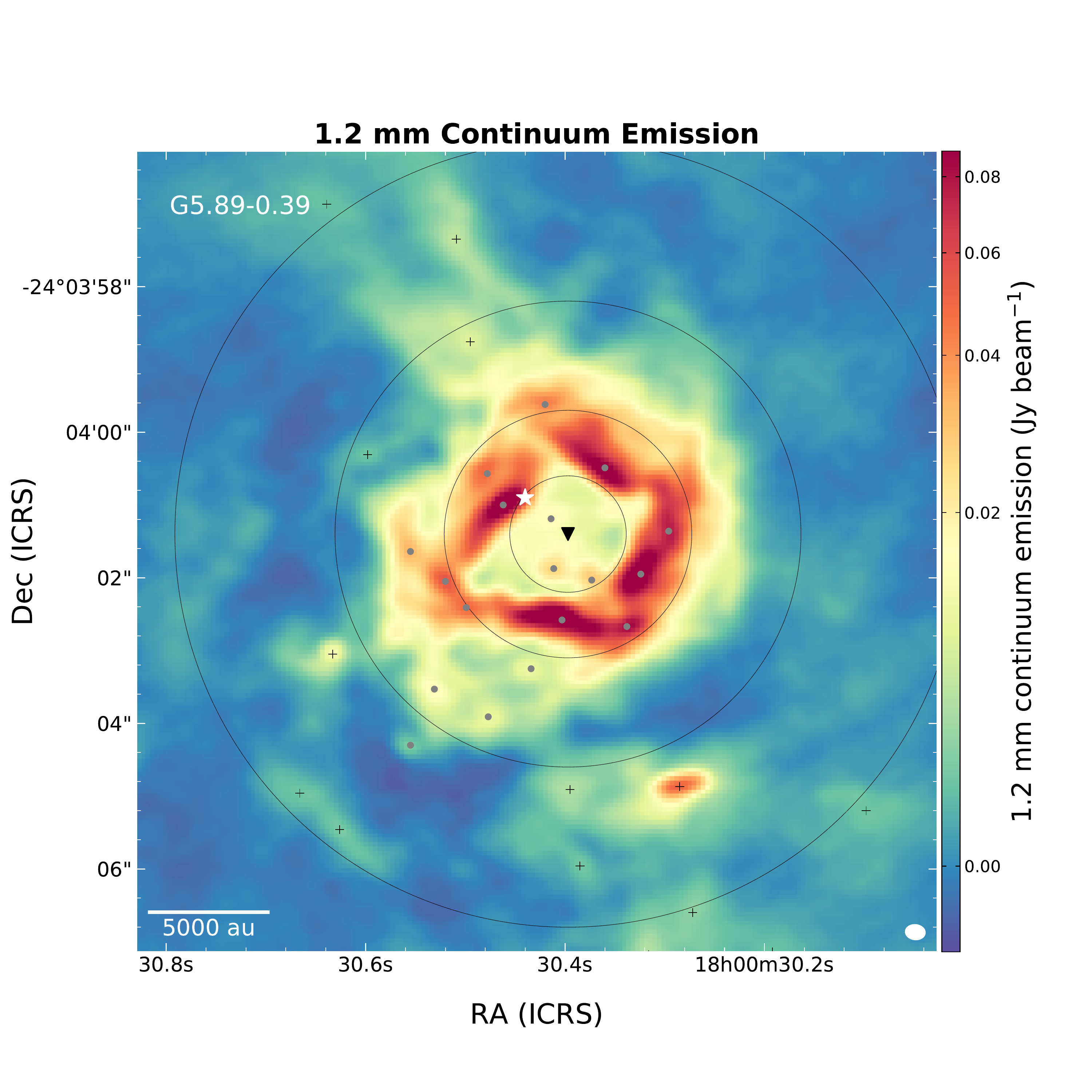}
\caption{Zoom in to the 1.2\,mm continuum emission of the Central Shell. A triangle marks the location we take as the center of the shell (ICRS 18:00:30.397, -24:04:01.4). The concentric circles of radii $0\farcs8$, $1\farcs7$, $3\farcs2$ and $5\farcs4$ demarcate the inner ionized region, the dusty belt, the ensemble of arches, and an outermost region with more faint arches and bow-shocks. The rest of the symbols are the same as in Figure \ref{f:overview}.}
\label{f:zoomin}
\end{center}
\end{figure*}

The high-angular resolution along with the image fidelity of the ALMA observations reveal a complex system of substructures toward the location of the Central Shell. Zooming into this region (Fig. \ref{f:zoomin}), we can distinguish condensations, arches and detailed morphology of the diffuse emission (with intensity levels of 1-4\mjy, which is between two and seven times 0.6\mjy, the local rms noise level in the outer shell zone), which reveal an intricate, structured and complex bubble of ionized gas and dust. 

The interior part of the shell-like structure (r$<0\farcs8$, with r measured from the center of the shell at ICRS\,18:00:30.397, -24:04:01.4) mainly contains ionized gas (average spectral index -0.08 consistent with the optically thin regime) plus three condensations (S11, S12, and S14). Its bounds, demarcated by a dusty belt of stronger emission, are irregular and curved, more elongated in the east-west than the north-south direction. The belt (seen as a shell-like structure within $0\farcs8<$r$<1\farcs7$) appears more four-sided than circular \citep[see also,][]{2003Feldt}. Moreover, the boundaries appear to be formed out of smaller and contiguous curved structures or arches. In spite of its irregularities, the boundary drawn by the ALMA 1.2\,mm emission does not show any apparent break in its continuity surrounding the inner zone. \cite{2008Hunter} reported a break in the souhtwest part of the belt (seen at 875$\mu$m and molecular emission) which is unseen in the ALMA data. 

Immediately surrounding the belt, the ALMA emission reveals the presence of arches, many of which apparently present their tips oriented radially outwards from the center of the shell structure. The ensemble of arches form a wispy pattern between radii $1\farcs7<$r$<3\farcs2$. In section \ref{s:dis_arcs} we make an attempt on identifying some of these arches, which include the millimeter envelopes (MM2, MM4, MM5, and MM7) and condensations (S1, S2, S3, S7, and S9) reported here, but the limited contrast and sensitivity of the image hampers conclusive delineation of the structures. 
A few of these arches, blurred in coarser angular resolution observations, have been reported in the past as loops of ionized and molecular gas in \target \citep{1998Acord,2000Argon,2008Hunter}, but the details of the new ALMA images allow one to disentangle their emission into substructures. In addition we note here that in the northeast and southwest directions, roughly coinciding with the large-scale dust filament orientations, there is a smaller amount of these arches.  

At larger distances from the center of the shell (between $3\farcs2$ and $5\farcs4$) there are more faint arches of dust (Fig. \ref{f:zoomin}). The tips of these arches also appear oriented outwards from the center of the shell, with position angles between $103\degr$ and $164\degr$ due southeast, and between $-77\degr$ and $-16\degr$ for their corresponding northwest counterparts.    

%\begin{figure*}
%\minipage{0.49\textwidth}
%  \includegraphics[width=\linewidth]{cont_zoom2.eps}
%\endminipage\hfill
%\minipage{0.49\textwidth}
%  \includegraphics[width=\linewidth]{cont_zoom.eps}
%\endminipage
%  \caption{\textbf{Left:} Zoom in to the 1.2\,mm continuum emission of the Central Shell. Several arch-like features can be discerned among the diffuse continuum emission. %Several curves (dotted black lines) are overlapped to aid the eye following the arches at larger distances from the center. All the archs are parts of ellipses with approximately the same center and dimensions. The center of these ellipses is marked by a magenta symbol close to the center of the shell at ICRS\,18:00:30.42, -24:04:01.6. 
%  \textbf{Right:} Another view of the Central Shell region.  %As in the left panel, the curves (black dotted lines) are patches of ellipses that share the same center (magenta symbol) at ICRS\.18:00:30.39, -24:04:01.4. 
%  Symbols are the same as in Figure \ref{f:overview}.}
%  \label{f:zoomin}
%\end{figure*}

%%%%%%%% Linear Pol
\subsection{1.2\,mm Linear Polarized Continuum Emission}
\label{s:pol}
\subsubsection{Overview}
\begin{figure*}
\minipage{0.51\linewidth}
  \includegraphics[width=\linewidth]{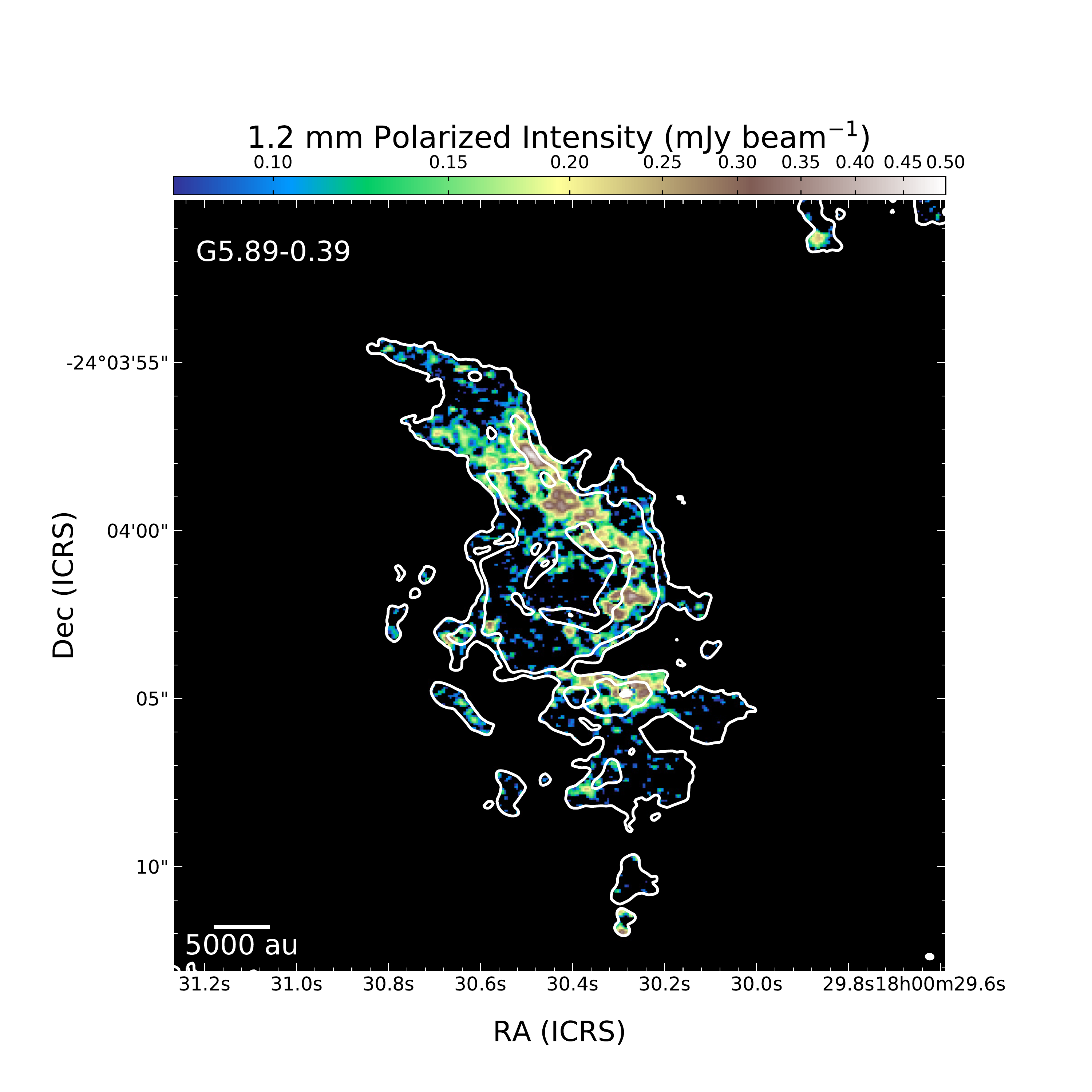}
\endminipage\hfill
\minipage{0.51\linewidth}
  \includegraphics[width=\linewidth]{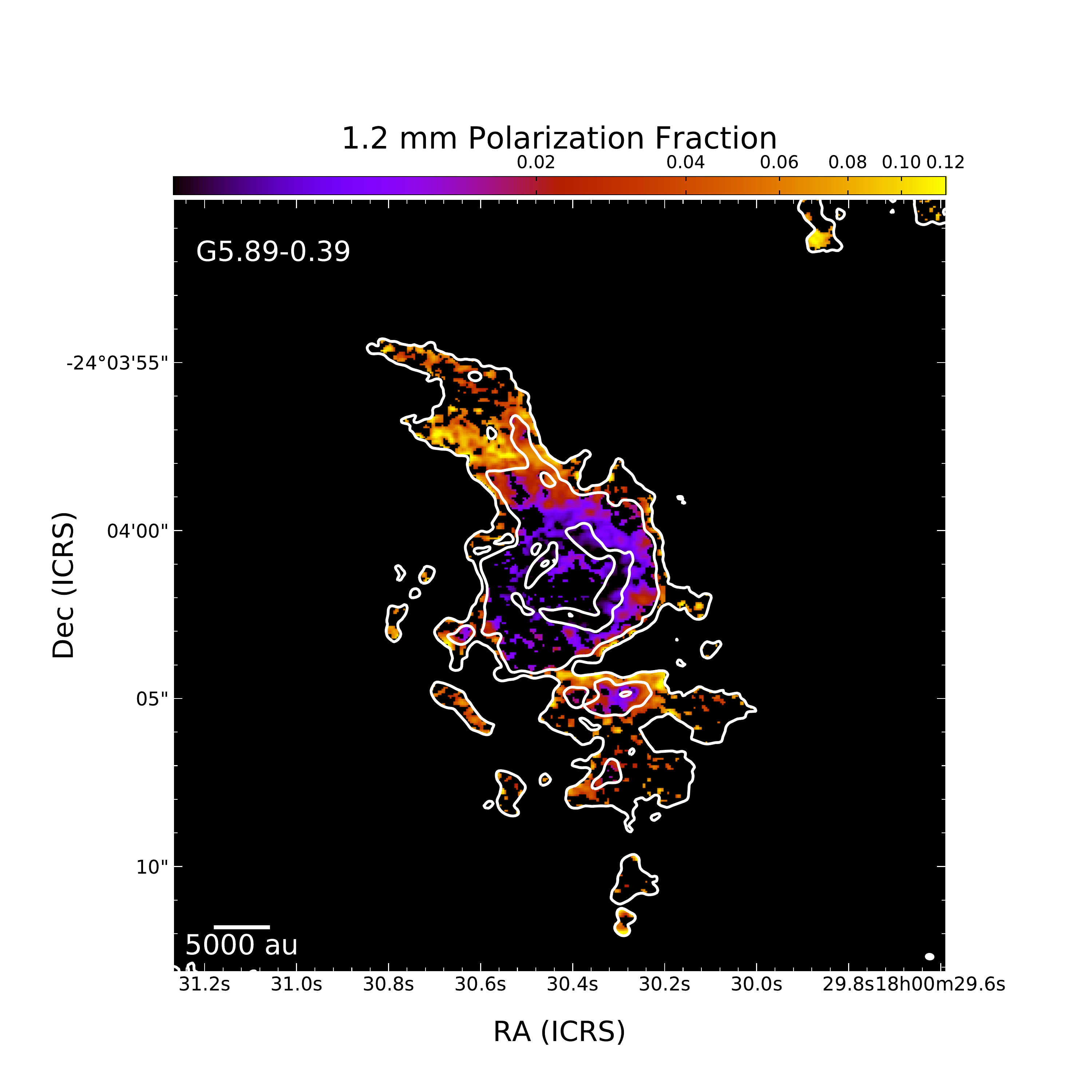}
\endminipage
  \caption{\textbf{Left:} Linearly polarized intensity overlapped with 1.2\,mm continuum emission contours at 1, 4, 40 and 100 times the rms noise level of 1.1\mjy. \ref{f:pa}. \textbf{Right:} Polarization fraction overlapped with continuum emission contours (same as left panel).}
  \label{f:pol}
\end{figure*}

The 1.2\,mm linear polarized intensity ($m_l$) spreads over the three characteristic regions in \target (left panel in Fig. \ref{f:pol}). It peaks at the MM15/SMA-S envelope ($m_l=$1.10\mjy) and shows a relatively strong southwest-northeast ridge along the North Filament (specially near MM9) and the western half of the Central Shell. We note that most of the polarized emission toward the Central Shell does not exactly coincide with the location of the brightest Stokes\,I emission from the belt. The area more polarized is instead shifted toward the exterior of the belt, as if it were surrounding the western half of the belt. Overall, the linearly polarized emission covers large areas of the North Filament and the north and west parts of the Central Shell. On the South Filament the polarized emission is more scarce, and it is mainly distributed in the surroundings of some millimeter envelopes such as MM15, MM13, and MM14 and MM16 at the southernmost tip. 

\begin{figure}
\centering
\includegraphics[angle=0, width=\columnwidth]{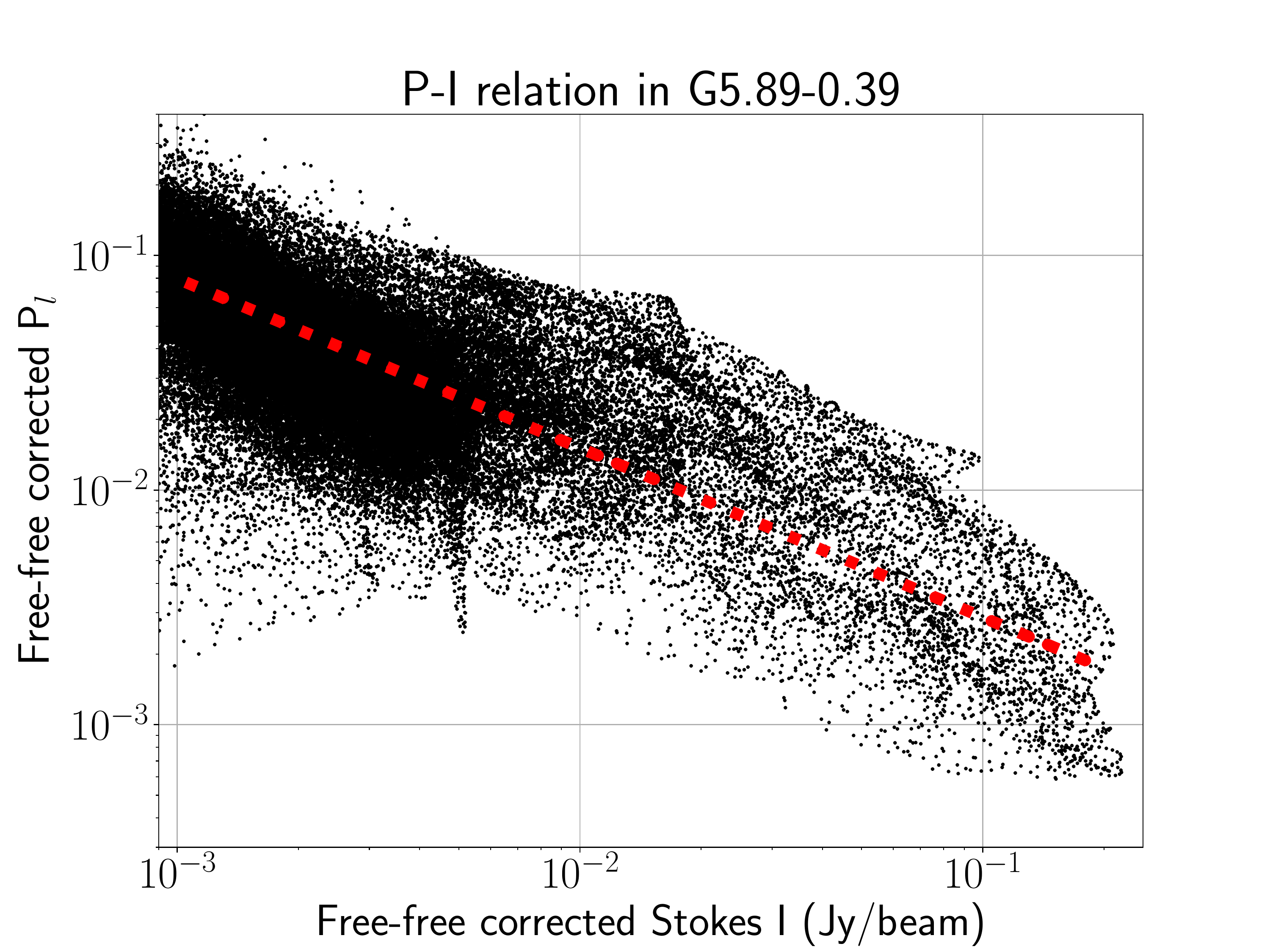}
\caption{Free-free corrected polarization fraction vs free-free corrected Stokes\,I intensity in \target. A power law trend with a slope of -0.72 was found when fitting the data with Stokes I over a $3\sigma$ threshold.}
\label{f:pi} 
\end{figure}

\begin{figure*}
\centering
\includegraphics[angle=0, width=\linewidth]{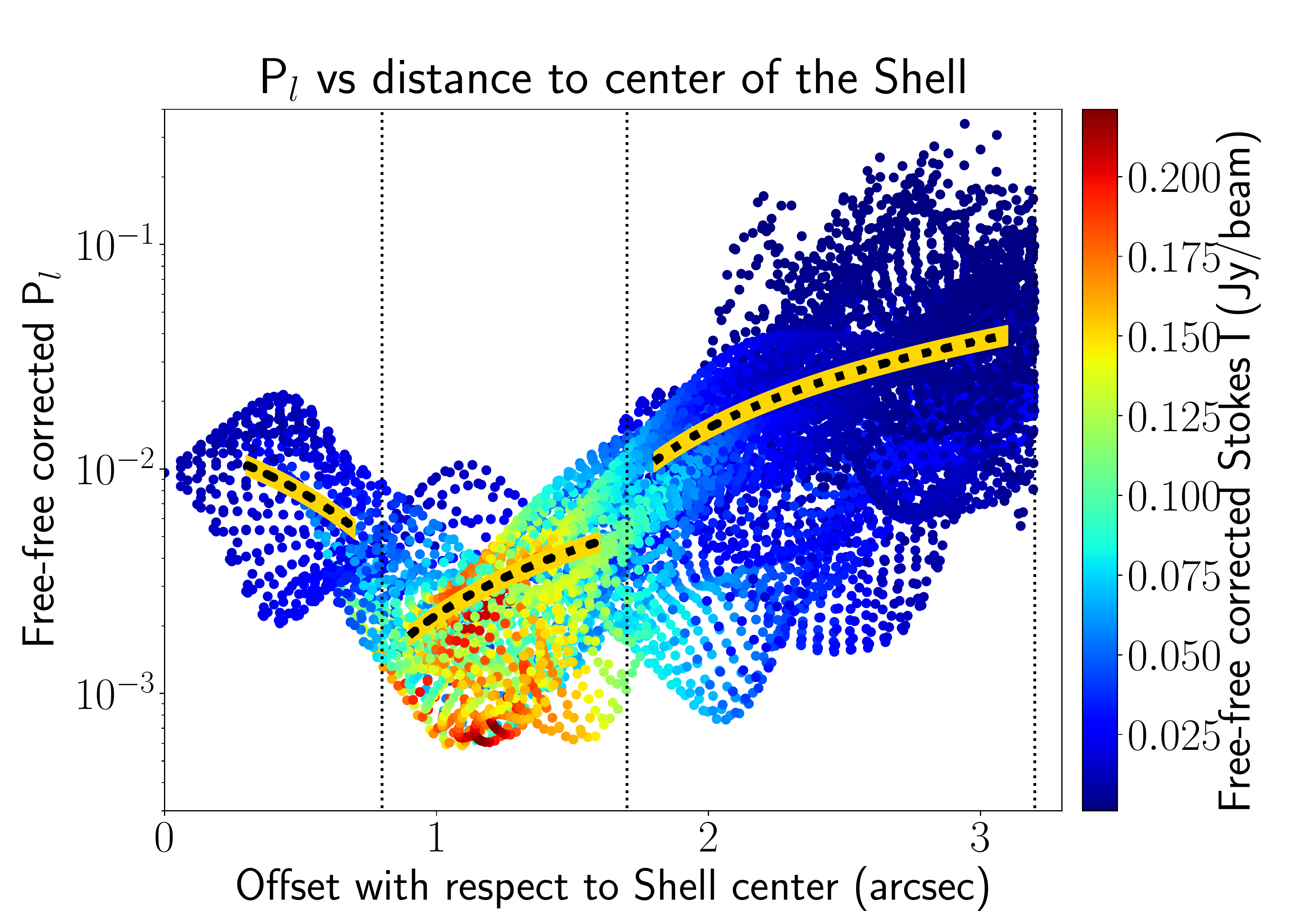}
\caption{Free-free corrected polarization fraction as a function of the distance from the shell center (ICRS\,18:00:30.397, -24:04:01.4). The color code shows stronger Stokes\,I emission in red and weaker emission in blue colored circles. Fitted power-law functions to the data of each zone of the Central Shell appear in dotted black curves highlighted in yellow bands.}
\label{f:pfrac_dist} 
\end{figure*}

The spatial distribution of the linear polarization fraction (p$_l$, right panel in Fig. \ref{f:pol}) shows an evident difference between the Central Shell region and the filaments. While p$_l$ is on average about 4.3-4.4\% in the North and South Filaments, with peaks well over 10\% at some specific locations (see also Fig. \ref{f:pi}), it drops down to 0.8\% on average in the Central Shell, mainly detected in its western part. The sudden drop in the polarization fraction toward the Central Shell cannot be explained totally by the contribution of the free-free emission at 1.2\,mm, since, as indicated in \cite{2009Tang}, free-free emission is completely unpolarized. That is, while the linear polarized emission is only due to dust, the Stokes\,I intensity has an additional contribution from free-free, thus contaminating (i.e., decreasing the value) the polarization fraction of the polarized dust emission (since the Stokes\,I emission produced by the dust would be lower, the dust polarization fraction should be larger than the measured). On average, we estimated that $\sim$40\% of the continuum emission at 1.2\,mm in the Central Shell is due to free-free; removing the effect of free-free emission in the Stokes\,I intensity, this would result in a factor of a few larger polarization fraction on average over this region. Hence, this correction does not change the fact that there is a sudden drop of p$_l$ over the Central Shell (a 0.8\% becomes a 2.1\%), it just indicates that the decrease is not as extreme. Figure \ref{f:pfrac_dist} shows the radial distribution of polarization fraction after removing the free-free unpolarized emission\footnote{We used the CASA programs IMREGRID, IMSMOOTH and IMMATH to produce the free-free corrected images. We first smoothed the ALMA data (both Stokes I and m$_l$ linear polarized intensity images) so that the images match the VLA angular resolution (Section \ref{s:vladata}). We then subtracted the extrapolated 1.2\,mm free-free contribution to the Stokes I image pixel by pixel. Finally we got a free-free corrected polarization fraction image, dividing the smoothed m$_l$ image by the free-free corrected Stokes I image. We use these data to build Figures \ref{f:pi} and \ref{f:pfrac_dist}.} from the Stokes\,I toward the Central Shell in more detail. It shows a gradual decrease in the polarization fraction at around the dusty belt position even accounting for free-free contamination (yellow bands). The corrected polarization fraction decreases from $\sim4.0$\% in the outer shell down to $\sim0.3$\% in the dusty belt. From the dusty belt inwards, the polarization fraction suddenly increases again to an average value of $\sim2.6$\%.

Figure \ref{f:pi} presents the relationship between the free-free corrected polarization fraction p$_l$ and the free-free corrected Stokes\,I intensity (also known as the P-I relationship) in the whole region. 
The data can be fitted by a power law
with a slope $s=-0.722\pm0.004$ (p$_l\propto I^{s}$), close to unity, indicating that the polarized intensity scales like the Stokes\,I intensity \citep[see also][]{2009Tang}. This is usually interpreted as the polarized intensity being produced in a surface layer of the molecular cloud, and attributed to grains aligned by radiation torques \citep[e.g.,][]{2019Pattle,2019Kwon}. However, the scatter at each Stokes\,I intensity (of up to one order of magnitude, e.g., from 0.7\% to 7\% at 10\mjy), still leaves the door open to different possibilities or trends at particular, smaller locations. The simple anti-correlation trend is probably not capturing all the essential physics of the polarized emission. Moreover, there are some departures from the anti-correlation trend at Stokes\,I intensities of 10-11\mjy and 50-100\mjy. These departures may indicate more complicated scenarios (see e.g., section \ref{s:mm15} below).
   %P-I relationship trends in the whole \target: \\
%Stokes I > 3.3 mJy \\
%m=-0.9810$\pm$0.0003 \\
%yes, this could even be emphasized more: The scatter is around one order of magnitude, and this very likely means that this simple anti-correlation is not capturing all the essential physics.

%Hay que analizar Pl vs I: al menos en MM10.
%Tang et al. 2009 dicen que:
%This is possibly due to a decreasing alignment
%efficiency in high-density regions, because the radiation torques
%are relatively ineffective (Lazarian \& Hoang 2007). It can
%also be due to the geometrical effects, such as differences
%in the viewing angles (Gon¸ calves et al. 2005), or due to the
%results from averaging over a more complicated underlying field
%morphology.
\begin{figure}
\centering
\includegraphics[angle=0, width=\columnwidth]{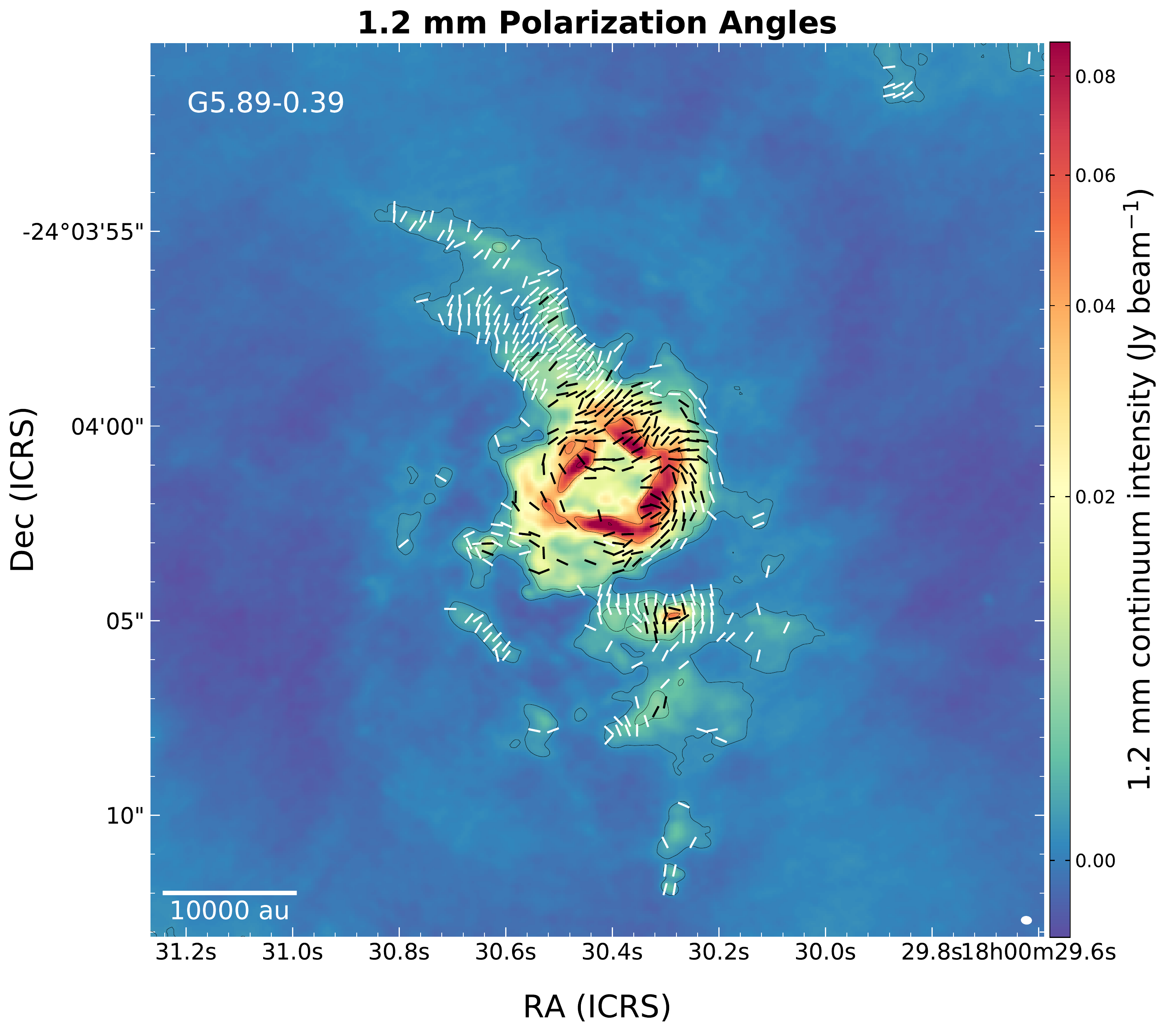}
\caption{EVPAs overlapped onto the Stokes\,I continuum emission towards \target. Black/White segments indicate the polarization percentage is below/above 2\% (no free-free correction applied to this image). A segment is displayed approximately per every beam, using data with a $2-\sigma$ cut in the debiasing process (same as in Fig. \ref{f:pol}). Contours are displayed at 1, 4, 40 and 100 times the rms noise level of 1.1\mjy.  
}
\label{f:pa}
\end{figure}

\begin{figure*}
\centering
\includegraphics[angle=0, width=\linewidth]{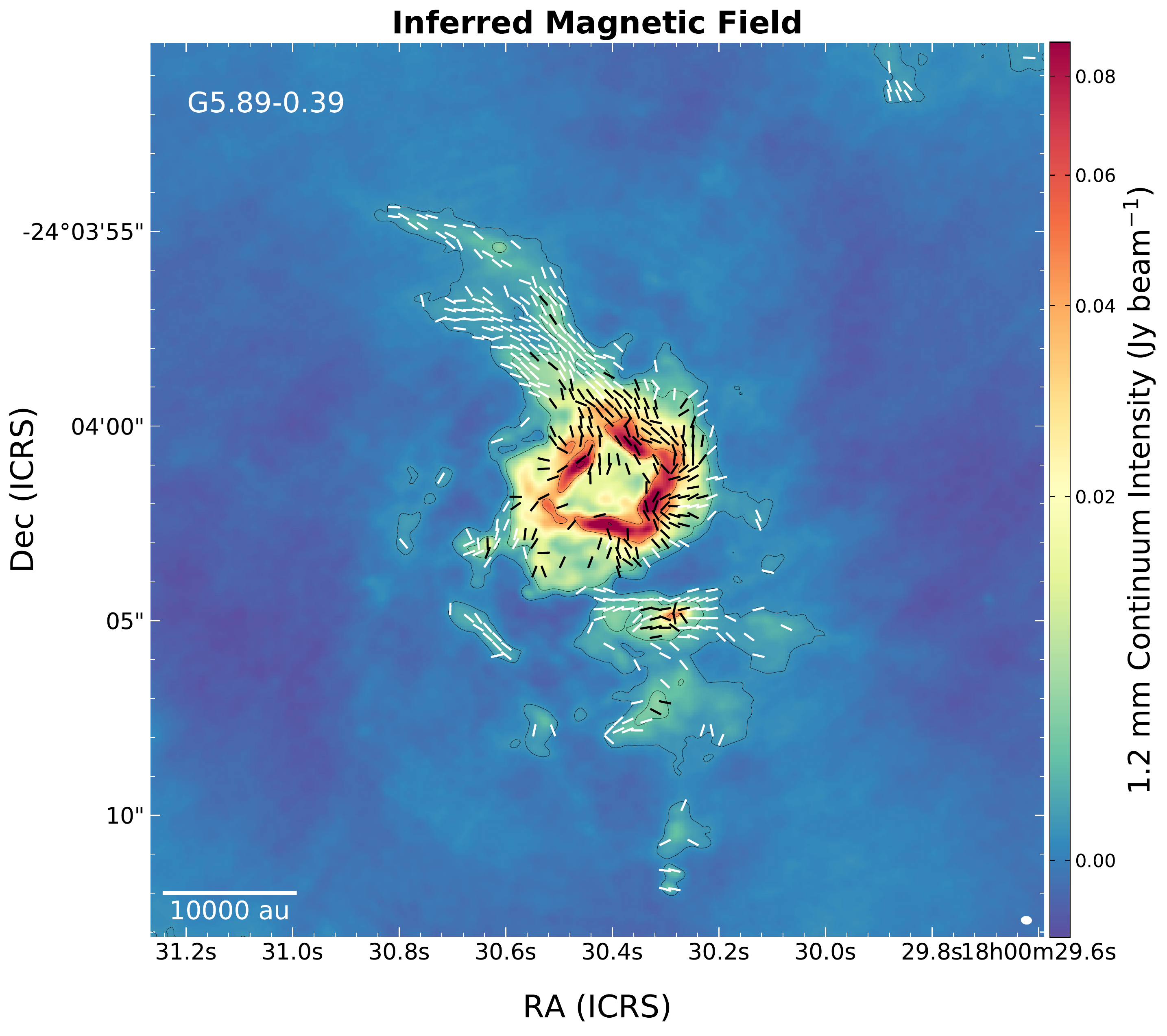}
\caption{Same as Figure 7, but with the EVPAs rotated by $90\degr$ to show the inferred magnetic field morphology.
%EVPAs rotated by $90\degr$ overlapped onto the Stokes I continuum emission towards \target. Black/White segments indicate the polarization fraction is below/above 2\%. Contours are displayed at 1, 4, 40 and 100 times the rms noise level of 0.0011\jy. 
}
\label{f:bpa} 
\end{figure*}

Figures \ref{f:pa} and \ref{f:bpa} show the spatial distributions of the EVPA and the inferred magnetic field orientations, respectively. To derive the magnetic field orientations, we  assume that these are orthogonal to the linear polarization orientations EVPA, as expected for grains aligned magnetically. 
%\citep[Davis-Greenstein mechanism][]{1951Davis}. 
From now on, we comment on the magnetic field orientations unless otherwise noted. 

\begin{figure}
\centering
\includegraphics[angle=0, width=1.05\columnwidth]{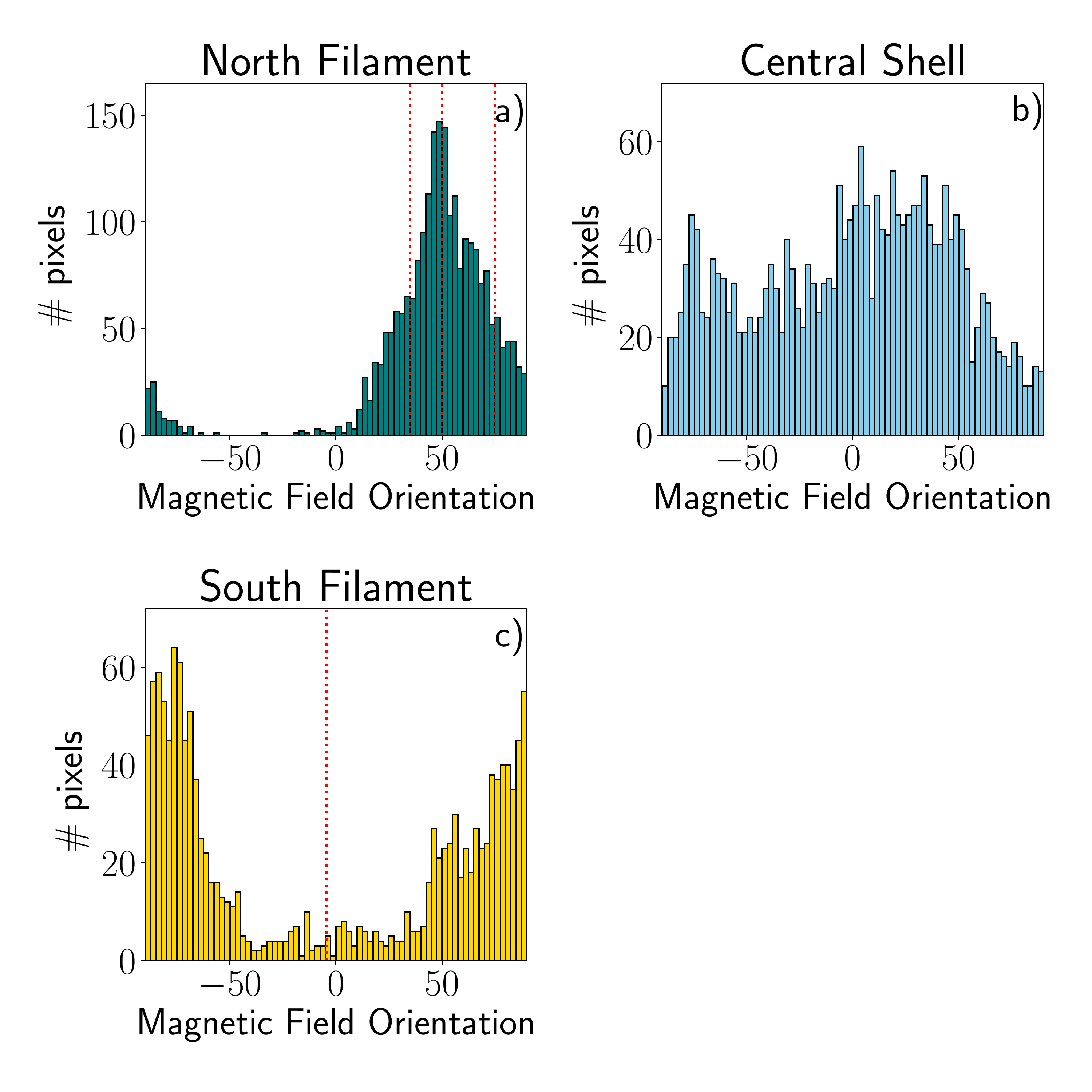}
\caption{Histogram of magnetic field orientations toward each one of the three main regions identified in Fig. \ref{f:overview}. \textbf{a)} The vertical red dotted lines mark the orientation of the North Filament (between $35\degr-50\degr$) and its northern tip ridge (at PA$=74\degr$). \textbf{b)} Central Shell histogram. \textbf{c)} The vertical red dotted line marks the average orientation of the South Filament at $-4.5\degr$. Both histograms for the northern and southern filaments (panels a and c) show clear single-peaked distributions.}
\label{f:bfields} 
\end{figure}

The North Filament shows a striking alignment of the magnetic field along its major axis (average PA of $54\degr$, Fig. \ref{f:bfields}), even accounting for the small turn of its northern tip, with a slightly different orientation than the rest of the filament. This can be easily seen in the histogram of panel a) in Figure \ref{f:bfields}, which shows a peak of magnetic field orientations close to the main filament orientation, with local angles between $35\degr$ and $50\degr$, and also in agreement with the northern tip orientation of $74\degr$. Similar magnetic field alignment was found for the filament-like streamers in IRAS\,16293-2422 region \citep{2018Sadavoy}

\begin{figure}
\centering
\includegraphics[angle=0, width=1.1\columnwidth]{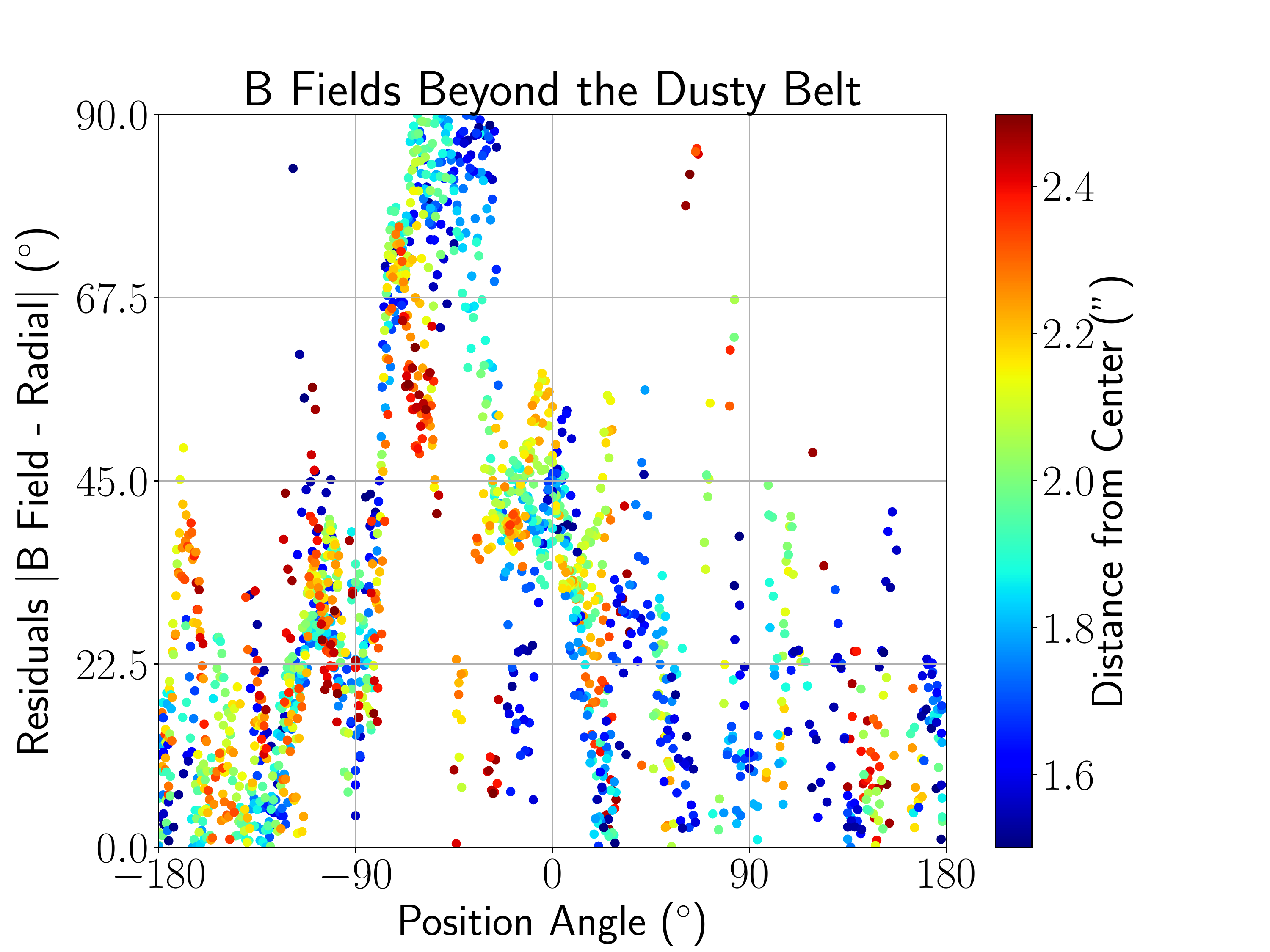}
\caption{Residuals (absolute value of the difference) between the magnetic field orientations and a radial field distribution plot as a function of the position angle measured with respect to the center of the Central Shell (ICRS\,18:00:30.397, -24:04:01.4), for pixels at radii between 1.5$\arcsec$ and $2\farcs5$. For the position angles, $0\degr$ points north and $+90\degr$ points east. Difference residuals close to zero indicate a good agreement with a radial pattern for the magnetic field. Note the excursion of the data away from the radial pattern in the northwest quadrant (around $-45\degr$), likely due to polarized emission related to the dusty belt. Excluding these data, the median and standard deviation of the residuals are $22\degr$ and $14\degr$, respectively.}
\label{f:pabradial} 
\end{figure}

At the location of the Central Shell, the magnetic field distribution seems more chaotic at first glance, but a more careful inspection leads us to see it is mostly radial beyond the belt of dust. Panel b) in Figure \ref{f:bfields} shows a more homogeneous distribution of the magnetic field orientations toward the whole HII region than in the filaments. This may be in accordance to the expectations for either radial or azimuthal distributions. However, when considering the magnetic field orientations in an annular region surrounding the dusty belt ($1\farcs5<r<2\farcs5$), the magnetic field shows a rough consistency with a radial pattern (Figure \ref{f:pabradial}). 

In the South Filament, the magnetic field segments follow a more simple distribution close to MM15, with an average orientation of $89\degr$, tracing the east--west elongation of the dust surrounding this envelope. This pattern is pinched at the very position of MM15 (see section \ref{s:mm15}). Otherwise, the polarization measurements are more scarce in the rest of this filament, with an average orientation of $88\degr$, which is almost perpendicular with the orientation of the filament itself (with local orientations typically of $-4.5\degr$). The magnetic fields around MM8 and MM21 change the orientation abruptly, while at MM14 and MM16, follow an east-west orientation, perpendicular to the filament.

\subsubsection{Polarization toward the MM15 Millimeter Envelope}
\label{s:mm15}
Table \ref{t:pol} summarizes the polarization properties toward some of the compact millimeter envelopes in which polarized emission is sufficiently detected (i.e., more than a beamsize area with polarized emission). An interesting case corresponds to the brightest of them, MM15, which shows a p$_l$ peak of 4.1\%. 

\begin{figure*}
\minipage{0.50\linewidth}
  \includegraphics[width=\linewidth]{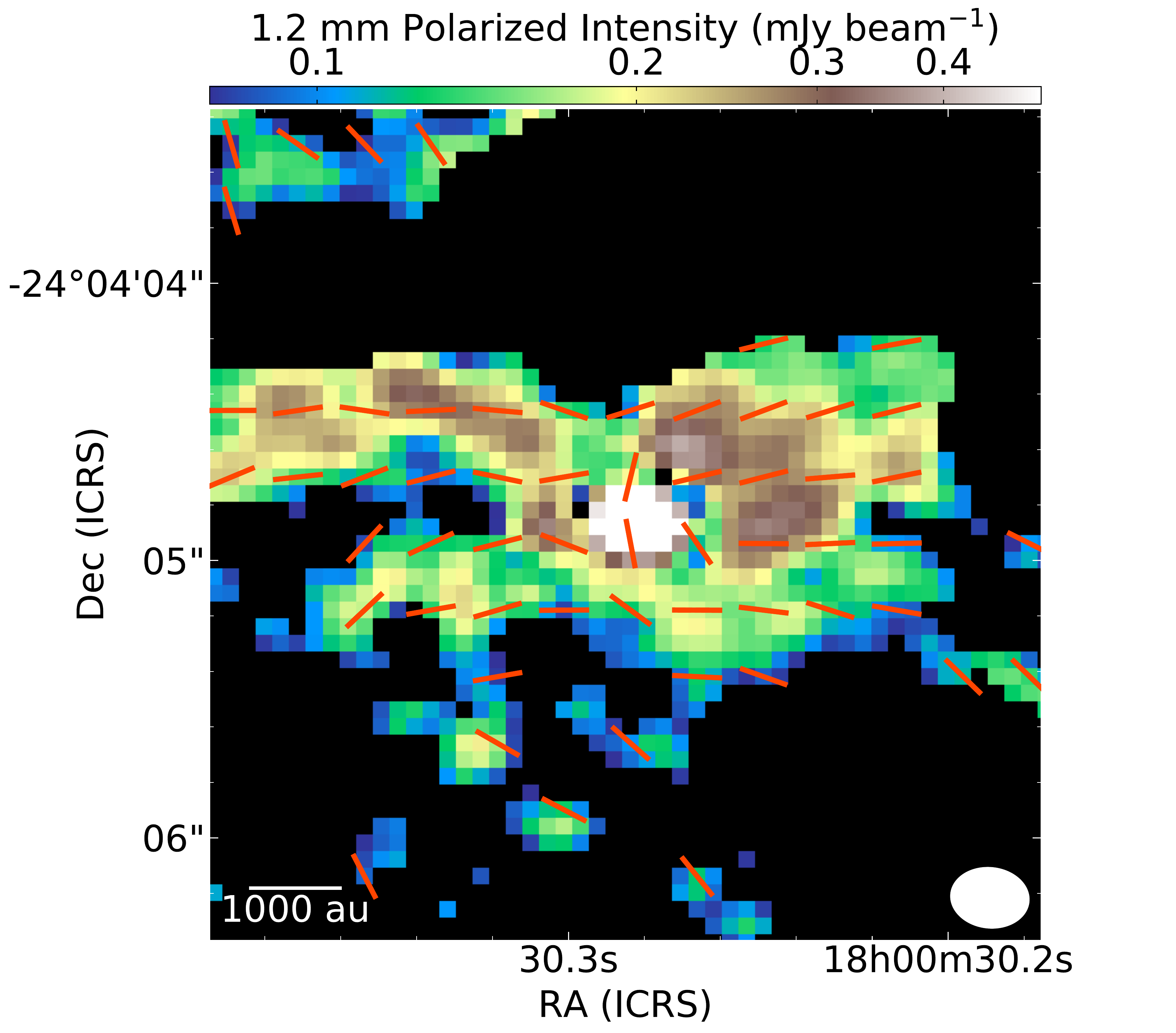}
\endminipage\hfill
\minipage{0.59\linewidth}
  \includegraphics[width=\linewidth]{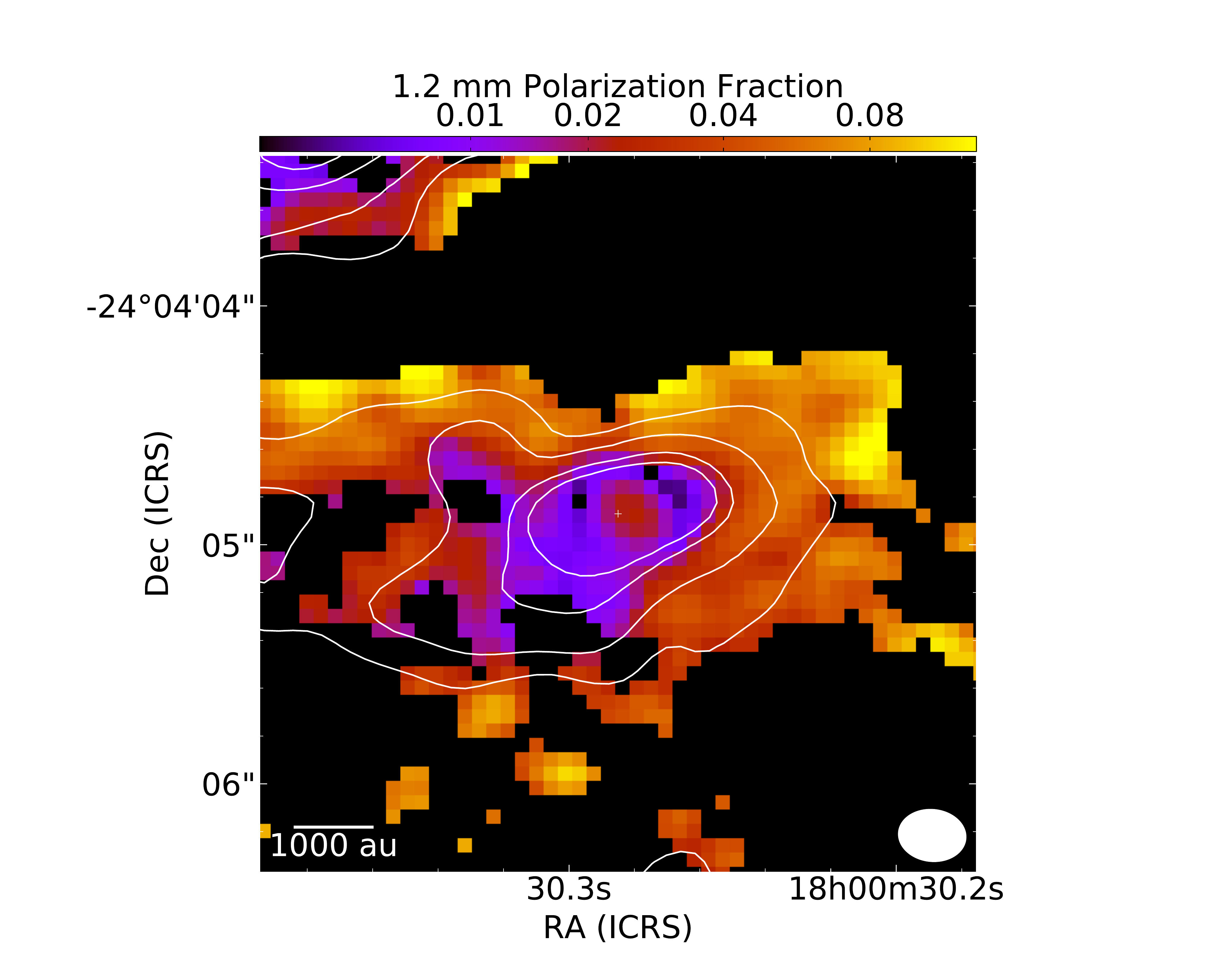}
\endminipage
  \caption{\textbf{Left:} Linearly polarized intensity overlapped with continuum emission contours toward MM15. Segments show the magnetic field orientations. \textbf{Right:} Polarization fraction overlapped with continuum emission contours toward the MM15 millimeter envelope (also known as SMA-S). Contours of the continuum emission in both panels are at 3, 5, 10 and 15 times the rms noise level (1.1\mjy). The synthesized beam is included at the bottom right corner. A white cross marks the position of the Stokes\,I peak emission.}
  \label{f:mm15_pol}
\end{figure*}

Figure \ref{f:mm15_pol} presents two panels with a zoom-in view of the MM15 surroundings. The left panel shows the centrally peaked distribution of the polarization intensity along with the magnetic field orientation. The right panel shows the spatial distribution of the polarization fraction with a local maximum (p$_l=$2.4\%) slightly offset from the continuum peak position $\sim0\farcs1$ west, surrounded by a ring of low polarization fraction (p$_l\simeq$1\%), which is likewise embedded in a region of emission with high polarization fraction (p$_l\simeq$7\%). 

\begin{deluxetable}{lccccccc}
\tablewidth{0pt}
\tablecolumns{7}
\tabletypesize{\scriptsize}
\tablecaption{Measurements on the polarization maps}
\tablehead{
\colhead{Zone} & \multicolumn{3}{c}{m$_l$ [\mjy]} &  \multicolumn{2}{c}{p$_l$ [\%]} &  \multicolumn{2}{c}{EVPA [$\degr$]}  \\
\colhead{}  & \colhead{F} & \colhead{P} & \colhead{Mean} & \colhead{P} & \colhead{Mean} & \colhead{Mean} & \colhead{$\sigma$}  } 
\startdata
%S7 	     & 0.92 & 0.35 & 0.18 & 0.7 & 0.3 & -54 & 16 & \\
%S8 	     & 0.10 & 0.29 & 0.19 & 0.06 & 0.03 & -71 & 8 & \\
%S9 	     & 0.88 & 0.37 & 0.19 & 0.8 & 0.3 & -10 & 59 & \\ 
%S10      & 0.20 & 0.24 & 0.14 & 0.4 & 0.2 & 14 & 64 & \\ 
MM1      & 0.06 & 0.12 & 0.10 & 6.4 & 3.6 & -61 & 14   \\
MM3      & 0.15 & 0.20 & 0.12 & 8.1 & 5.0 & -20 & 17  \\
MM4      & 0.30 & 0.18 & 0.12 & 3.7 & 1.8 & 86 & 78  \\
MM5      & 0.28 & 0.17 & 0.12 & 6.3 & 4.4 & -11 & 11  \\
MM9      & 0.40 & 0.39 & 0.21 & 5.7 & 3.2 & -53 & 6 \\
MM10     & 0.27 & 0.27 & 0.20 & 3.0 & 2.1 & -62 & 7 \\
MM13     & 0.23 & 0.16 & 0.11 & 3.3 & 2.3 & -21 & 15  \\
MM14     & 0.32 & 0.36 & 0.20 & 18.6 & 8.6 & -11 & 5  \\
MM15     & 1.58 & 1.10 & 0.32 & 4.1 & 1.4 & -66 & 46  \\
MM16     & 0.51 & 0.36 & 0.18 & 14.1 & 7.7 & -28 & 11  \\
MM23     & 0.58 & 0.23 & 0.16 & 12.6 & 9.7 & -67 & 6  \\
%inner    & 0.87 & 0.21 & 0.12 & 1.6 & 0.9 & -25 & 75  & 0.0027 0.012 \\
%belt     & 13.0 & 0.42 & 0.18 & 2.5 & 0.6 & -29 & 48  & 0.0001 0.0001 \\
%exterior & 7.57 & 0.33 & 0.13 & 15.0 & 3.7 & -11 & 48 & 0.0046 0.010 \\
\enddata 
\tablecomments{m$_l$ is the linear polarized intensity in \mjy; p$_l$ is the degree of linear polarization (in\,\%); EVPA is the electric vector position angle (in\,$\degr$). 
F designates the integrated flux, P the peak intensity value, Mean is a measure of the mean value and $\sigma$ is the standard deviation taken over the pixels inside manually defined boxes for every source. Angle values were extracted using directly average values on Stokes Q and U images.}
%\tablenotetext{(a)}{Deconvolved semimajor and semiminor axis of the ring-shaped disk.}
\label{t:pol}
\end{deluxetable}

In general the magnetic field distribution toward MM15 roughly follows the east--west elongation of the dust emission. However, close to the continuum peak, the orientation of the magnetic field segments have a dramatic $\sim90\degr$ jump (Fig. \ref{f:mm15_pol}). 

\begin{figure}
\centering
\includegraphics[angle=0, width=1.1\columnwidth]{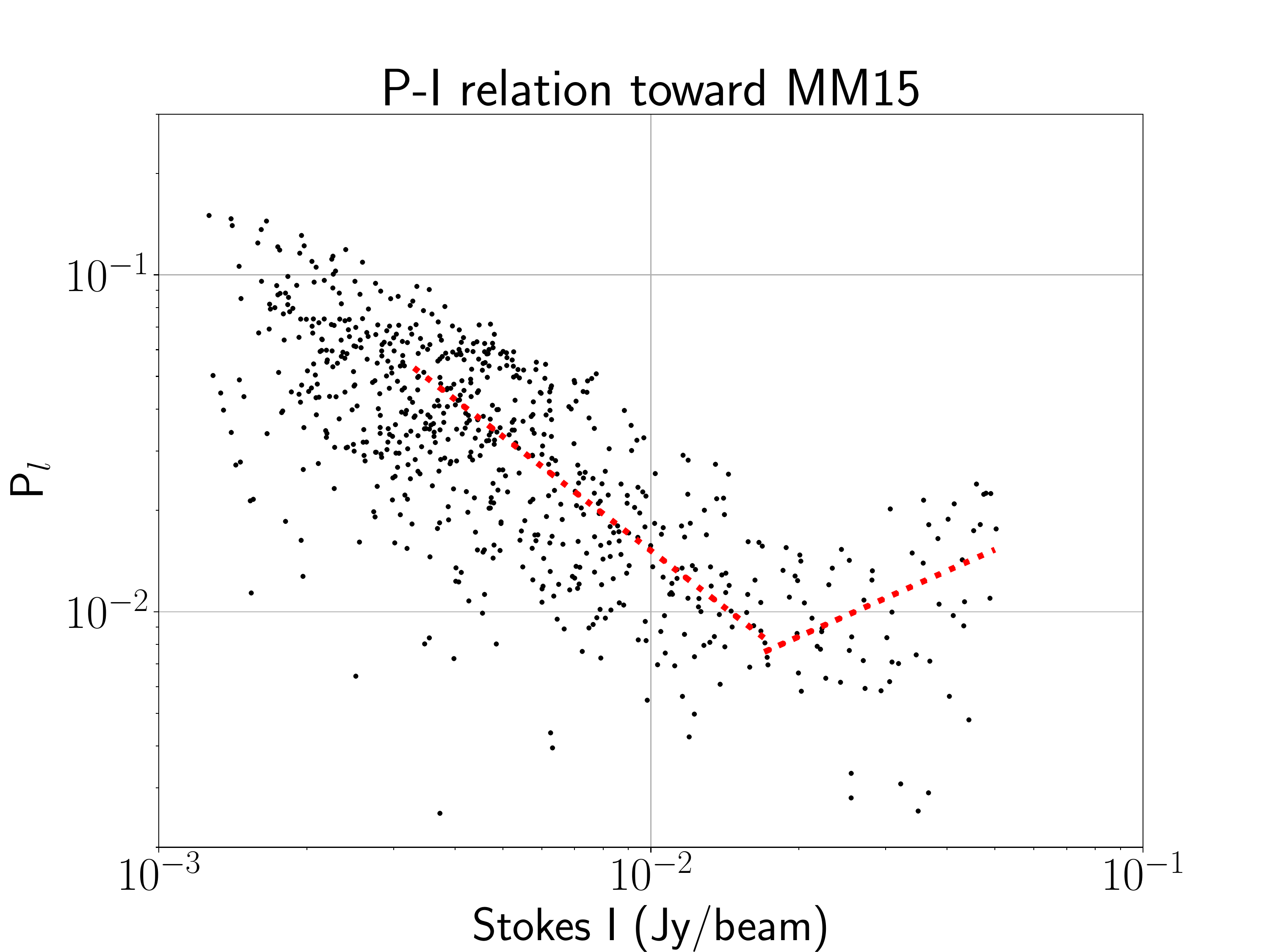}
\caption{P-I relation towards the MM15 millimeter envelope. Two linear trends are plotted for Stokes\,I intervals between 3.3 and 17\mjy and $>17$\mjy, respectively.}
\label{f:pi_mm15} 
\end{figure}

Figure \ref{f:pi_mm15} presents the P-I relationship in the neighborhood of MM15. The data can be clearly divided into two regimes. We fit the data with Stokes\,I emission between 3.3\mjy and 17\mjy with a power-law with a slope $s=-1.1\pm0.1$. For the data over 17\mjy, despite the larger scatter, a power-law fit gives a slope of $s=0.6\pm0.2$. A negative/positive slope in the P-I plot indicates lower/higher efficiency in polarization at higher dust densities.

%P-I relationship trends: \\
%Stokes I between 1.1 and 17 mJy \\
%m=-1.127$\pm$0.097 \\
%Stokes I between 17 and ~50 mJy \\
%m=0.648$\pm$0.23 \\

\section{Discussion} 
\label{s:discussion}
\subsection{General Scenario for the Star Formation in \target}
The presented 1.2\,mm ALMA continuum observations toward \target revealed a filamentary structure about $18\arcsec$ long ($\sim0.25$\,pc), interrupted by the well-known ionized shell-like structure at its center which hosts an extremely energetic explosive outflow \citep{2019Zapata,2020Zapata}. We propose here a scenario in which the North and South Filaments were part of the same molecular cloud structure. They were possibly part of a single curved filamentary cloud, hosting a small cluster of intermediate- and high-mass protostars at the position where we now see the center of the Central Shell. This central cluster may have formed an unstable multiple system which finally unleashed an explosive dispersal event \citep[as it seems to have occurred in the Orion BN/KL system][]{2005Bally,2005Rodriguez}, driving gas and dust debris outward quasi-isotropically at high velocities. 
After the explosive dispersal event, the HII region (which may or may not have existed before the blast) includes hot ionized \ion{H}{2} gas mixed with warm dust; the latter is mainly concentrated in the aforementioned $\sim4500$\,au radius projected belt-like structure. Possibly embedded in this dusty belt is the only stellar object confirmed to date: the O5 Feldt's star, probably the main ionizing source. However, there also is evidence for another star as identified by \cite{2006Puga}, using a velocity gradient attributed to its outflow. The outflows from the proto-cluster members, the expanding UCHII region, and the explosive dispersal event may have shaped the environment until getting the present shell-like appearance. Furthermore, there is reasonable evidence that the ejected gas and debris are probably reaching some of the more nearby star-formation centers, such as MM15 \citep[e.g., see SiO molecular images in][]{2020Zapata}.
 
In more remote places of the molecular cloud/filament, the on-going process of filament fragmentation is probably unaffected by the explosive event, as the shockwave has not yet reached these areas. In these locations, several envelopes have already gravitationally condensed, creating protostars revealed by the millimeter emission from dust and the presence of collimated outflows (such as MM1 and MM6, which are associated with colimated bipolar outflows seen in $^{12}CO$, which we will show in a forthcoming publication). While the fragmentation process has not disrupted the overall structure of the North Filament (note however that it diverges into two sub-filaments), the South Filament is more patchy, with three local star-formation centers associated with the millimeter envelopes MM13, MM15, and MM17. Nonetheless, the filamentary nature of the South Filament can be perceived in the almost north-south alignment of several of its envelopes, while still preserving the original large-scale structure.

%\textbf{Is the following necessary/appropriate?}
%\textcolor{red}{An even closer inspection to the ALMA continuum observations reveal a very low level emission structure like a possible large-scale semi-circumference southeast of the Central Shell, and some possible striations west of the South Filament, but we did not discuss them here, since the sensitivity is not enough to clearly distinguish them well over the noise threshold.}

\subsection{Orientation of the Polarized Emission in the Central Shell}
%As we said before, the Central Shell is probably the result of an explosive outflow event occurred about 1000 years ago \citep{2020Zapata}. 
The main morphology distribution of the linear polarization EVPAs (i.e., not rotated to show the inferred B-field) in the Central Shell, the azimuthal polarization pattern beyond the dusty belt (Fig. \ref{f:pa}), may be explained by emitting dust grains aligned with the radial magnetic field \citep{1994Lazarian,2007Cho,2007Lazarian}. We assume this explanation in Section \ref{s:results}, but we discuss here whether this distribution could possibly also be explained by the alignment via radiative flux \citep[e.g.,][]{2017Tazaki} coming from the center of the explosion as a locally anisotropic flow of photons, or even by a process of self-scattering provided by anisotropic thermal dust emission \citep{2015Kataoka,2016Yang}. These latter two alignment mechanisms need dust grains with sizes about $\lambda/2\pi$ ($\sim0.2$\,mm), which seems implausible, for typical interstellar dust grains are micron-sized. Additionally, a scenario with grains aligned mechanically by an isotropically expanding blast has to be rejected in principle, because the grains will be oriented radially and will produce a radial polarization pattern \citep{1952Gold}, opposite to the azimuthal encountered in the ALMA images. However, the alignment via mechanical torques \citep[e.g.,][]{2018Hoang} may explain the observations of the magnetic field is radial. In addition, a similar azimuthal pattern has been observed up to $\sim4500$\,au from the center of the explosive dispersal outflow in Orion\,BN/KL \citep[e.g.,][]{2020Cortes}. The mechanical alignment of grains was claimed to explain sudden changes in the orientation of the polarization at larger scales in this region \citep{1998Rao}, but \cite{2010Tang} and \cite{2020Cortes} favored the magnetic alignment scenario to explain the millimeter polarized emission. Furthermore, they estimate that the explosive outflow may be energetic enough to drag the magnetic field lines, and arrange them into the observed radial pattern. Assuming both explosive events produce similar outcomes, we also favor the scenario of magnetically aligned grains in the Central Shell of \target as the main contributor to the millimeter polarized emission. 

The present ALMA observations toward \target show polarized emission at the belt of dust emission but primarily outside it, in the less dense parts of the Central Shell (Fig. \ref{f:bpa}). The polarized emission is more likely found in the western hemisphere of the shell. Along the dusty belt, the orientation of the magnetic field segments is irregular although many appear to be azimuthal with respect to the center of the shell, especially at the location of the millimeter condensations (Fig. \ref{f:bpa}). Beyond the dusty belt, the orientation of the magnetic field is predominantly radial (Figs. \ref{f:pa} and \ref{f:pabradial}). 
The polarization percentages decrease progressively inwards down to the inner boundary of the dusty belt (Fig. \ref{f:pfrac_dist}). From that point to the the center of the shell, the polarization fraction increases again. This can be another manifestation of the \lq\lq polarization hole\rq\rq, where the polarization fraction gets lower as Stokes I emission gets higher due to preferential alignment at a surface layer of the dusty belt or due to depolarization effects (see also section \ref{s:lptmm15}). In this case, the sudden change on the magnetic field orientation at the outer edge of the dusty belt may be connected to the polarization fraction drop (e.g., due to tangled up magnetic fields), which can raise a question on whether the fields lines are being reshaped due to the ionization and/or the expansion of the shell. What is clear is that a radial distribution does not correspond to the idealized models of expansion of bubbles through uniform magnetic fields \citep[e.g.,][]{1991Ferriere}.
One possible scenario is that during the first stages of the explosion, the compression on the dusty belt made the fields azimuthal. In more evolved stages, at the position of the condensations in the belt (zones of maximum compression), the magnetic field resistance against the expanding wave is maximal, acting as stagnation points where the gas ram pressure and the thermal pressure is balanced by the magnetic field tension. If magnetic fields are flux-frozen to the gas, they will be dragged along with the outgoing motions creating an outwards radial magnetic field pattern.

In any case, finding a prevailing radial magnetic field accompanying the explosive dispersal toward \target \citep{2019Zapata,2020Zapata}, makes two such systems with the same magnetic field signpost \citep[the other being Orion BN/KL][]{2020Cortes}, which are unique in detecting the relation between the feedback event and its fields. This signpost \citep[more difficult to detect in low-angular resolution observations and/or at older evolutionary stages][]{2016Planck34,2017Soler} should be tested in other explosive dispersal candidates as it could be an independent way of corroborating the existence of these special protostellar outflows.  

\subsection{Analogous Magnetic Field Morphology in Supernova Remnants}
In \target, the shell comprises both ionized gas and dust and it is thought to expand at $\sim35$\kms \citep{1996Afflerbach,1998Acord}. This is reminiscent of the shells created by planetary nebula, supernova remnants \citep{1991Gomez} or even superbubbles \citep{1991Ferriere}, whereas in this case, the shell is at the center of an interstellar medium filament showing an incipient star-formation process, and thus it is at a very young stage. Furthermore, we note here that the physics of the expansion of an HII region is different from those of a supernova remnant, and the primary source of radio emission in supernova remnants is synchrotron emission, characterized by a negative spectral index and an expected large degree of linear polarization at centimeter wavelengths. Although the basic model of an expanding bubble may explain the overall physics in \target, the escape of hot gas and winds are significant effects \citep{2020Geen}, not clearly stablished in this region. None of these have been observed toward \target \citep[][]{1996Afflerbach,2004Sollins,2008Hunter,2009Tang}. Nonetheless, we find similarities between the distribution of the magnetic fields in \target with those found in supernova remnants, which may be interesting to explore \citep[such as Cas\,A, see][]{2009Dunne}. 

%None of these have been observed toward \target, since the Spectral Energy Distribution (SED) of the shell is consistent with that of an \ion{H}{2} region \citep[][]{1996Afflerbach,2004Sollins,2008Hunter}. Moreover, \cite{2009Tang} reported zero polarized light at radio wavelengths. Nonetheless, we find similarities between the distribution of the magnetic fields in \target with those found in supernova remnants \citep[such as Cas\,A, see][]{2009Dunne}. 

In their review, \cite{2015Dubner} comment that the magnetic field orientation in supernova remnants may depend on their age. Young shell-type remnants present a radial distribution of magnetic fields, whereas in older remnants the distribution is predominantly parallel to the front shock or it is tangled \citep{1987Milne}. The origin of the radial distribution at the interface between the expanding front of young remnants and the swept-up environment, could be produced with the amplification of the fields by the stretching of Rayleigh-Taylor instabilities in the mixing layer \citep[e.g.,][]{1996Jun}, but this is still a matter of debate \citep[see e.g.][for an alternative model]{2017West}.

Nevertheless, the conclusion is that synchrotron polarized emission from supernova remnants is oriented orthogonal to the magnetic fields, as with the magnetically aligned grains case. If the polarization at millimeter wavelengths is produced by a mechanism related to the magnetic fields (grains alignment or synchrotron emission), the orientation of the EVPAs would be expected to be perpendicular to the field lines. Hence, the magnetic field lines would be predominantly radial, as observed in \target. The possibility that the explosive event in \target somehow produced particles accelerated up-to relativistic velocities and created a small population of synchrotronic plasma seems unlikely at first glance. However, this scenario would be supported by the detection of high-energy sources \citep[from X-rays to $\gamma$-rays, see e.g.][for a summary]{2015Gusdorf} matching the position of \target \citep{2016Hampton}, although the usual explanation for the HESS source is that it is produced by the interactions of cosmic rays accelerated in the northern W28A2 supernova remnant with the dense gas of the star-formation region. 

More detailed observational work is needed (multiwavelength deeper images) to disentangle the different contributions to the emission (thermal dust, thermal free-free and possible non-thermal synchrotron emission) to extract more information about the physical conditions of the gas and dust in the region. This will help to obtain a better informed interpretation of the explosive event and its interaction with the previous outflow ejecta and the possible pre-existing \ion{H}{2} region.

\subsection{Orientation of the Polarized Emission in the Filaments}
In most regions of the insterstellar medium, it is usually assumed that dust grains are aligned with their long axis perpendicular to the magnetic field lines through the effect of radiative torques \citep[RATs,][]{1996Draine,2007Lazarian,2008Lazarian,2008Hoang,2009Hoang,2016Hoang,2019Lazarian}. Here, we presume this is the situation in most of the North and South Filaments.  

Observations with the Planck satellite \citep{2016PlanckCollaboration} have revealed that elongations of gas with column densities below $5\times10^{21}$\cmd in the diffuse interstellar medium are predominantly aligned with their associated magnetic fields, while more dense filaments (column densities over $5\times10^{21}$\cmd) are found to be mostly perpendicular to the magnetic fields \citep[][and references therein]{2019Hennebelle}. In \target, the ALMA polarization observations show that the elongation axis of the North Filament is mostly aligned with the magnetic fields, and the South filament is perpendicular to them\footnote{Throughout this paper we presumed the orientation of the southern filament to be along a north-south direction due to the clear orientation of the northern filament, which suggests a further connection south of the Central Shell. Nevertheless, some of the southern more dispersed continuum structures actually have an east-west orientation prevailing at smaller-scales. The magnetic field direction may be more related to this smaller-scales orientations.}.
The average column density for both filaments is similarly high (even taking into account their lower derived masses, we obtain values higher than $\sim10^{24}$\cmd), lying both in the high-density regime in which the filaments are expected to be perpendicular to the magnetic fields. 
The North Filament does not comply with this expectation. One possible explanation is that the magnetic fields, although orthogonal to the filament, are contained in a plane perpendicular to the plane of the sky, and therefore, we would see their projection as aligned with the filament orientation. \cite{2020Doi} estimate that the probability of this special 3D-spatial configuration for filaments with orthogonal magnetic fields is 14\% (allowing a 30$\degr$ separation from the perpendicularity of the plane containing the magnetic fields and the plane of the sky). Hence, this configuration has low probability. Another explanation may come from the theoretical predictions for weakly magnetized filaments. For these, turbulent motions and gravity largely dominates the energy budget. \cite{2013Soler} have found that for weakly magnetized filaments, even in the case of large densities, the magnetic field may be predominantly aligned with the filament \citep[see also,][]{2019Hennebelle}. For \target, this may be the case for the North Filament, while the magnetic fields may play a different, more active role, in the energy budget of the South Filament. \cite{2017SolerH} have found in addition, that both, aligned and perpendicular configurations between magnetic field and filament orientation, may be preferred configurations in numerical simulations. The perpendicular configuration may only be seen in converging flows threaded by more prominent magnetic fields. In addition, given an initial magnetic field strength and high enough density, switching between both configurations may be triggered by compressive motions.

In any case, \target is a challenge for both hypotheses, since the North and South filaments may be possibly part of the same structure (apparent continuous north-south morphology) and present two separate magnetic field distributions. 
%We note here that the position angle measured from the tip of the South Filament to that of the North Filament is $\sim28\degr$. Interestingly, the whole structure is almost parallel with the plane of the Milky Way (PA$\simeq30\degr$). 
The reason behind the magnetic field differences between the North and South Filaments is thus unclear. It could be related to intrinsic orientation (projected geometry of the filaments) or strength changes in the magnetic field. The latter could likewise be related to differences in the star-formation evolutionary stage of the envelopes embedded in both filaments. For instance, the South Filament could be more advanced in terms of gravitational fragmentation and collapse, as indicated by the many local centers of star formation. Fig. \ref{f:bfields} shows that close to the main millimeter envelopes, the magnetic field lines are pinched and present a lower polarization fraction. This could indicate that the field lines geometry is dominated by gravitational processes, hence explaining their different orientations with respect to those in the North Filament.

An alternative scenario may involve a strong interaction of the explosive outflow with the South Filament and a milder interaction with the North Filament, which maybe somehow shielded because of the geometry and dust density distribution of the system. There is evidence showing the probable impact of some explosive outflowing filaments against the dust pocket surrounding MM15 \citep{2020Zapata}. The north side of this pocket shows magnetic field lines mainly oriented in the east-west direction, whereas there are a few segments in the southern side  aligned in the northeast-southwest direction. In this scenario, the east-west orientation of the field lines (roughly orthogonal to a vector from the shell center) may be the product of compression due to shocks at the northern border of the dust pocket, as in MM2 and MM5. This may also be the case for the surrounding dust of other envelopes such as MM13 and MM22, although more evidence should be gathered. Regardless of this alternative scenario, the fact that we stress in this section is that in the filamentary structure of \target the magnetic fields experience abrupt orientation changes between the North and the South filaments.

%2) South Filament: uniform magnetic field aligned with the elongation of SMA-S. Fields are pinched probably due to gravitational free-fall of the material surrounding the core. When a filament has already undergone gravitational fragmentation around some overdensities, the magnetic field anchored to the ions starts to follow the dynamical motion of the free-fall and is probably dominated by gravity.  

\subsection{Linear Polarization Toward MM15}\label{s:lptmm15}
At $>2000$\,au spatial scales, the magnetic field spatial distribution around MM15 seems to follow an uniform  orientation roughly in the east-west direction (Fig. \ref{f:mm15_pol}). This direction is perpendicular to the South Filament orientation, but it is also locally perpendicular to the explosive outflow from the Central Shell. Hence, it is difficult to decide whether the field orientation is due to the perpendicular accretion flows channeled into the South filament or a process of compression due to the explosive dispersal outflow.
  
At $<2000$\,au spatial scales, toward the center of the MM15 millimeter envelope, the magnetic field orientation jumps from the east-west to a more north-south direction. This is reminiscent of the expectations for a magnetically regulated collapse \citep[e.g.,][]{2006Girart,2009Girart,2013Stephens,2018Maury,2019Kwon}, where the initially uniform distributed magnetic field lines are mainly distributed along the rotation axis of the system, but dragged and pinched as the collapse advances. The MM15 envelope is elongated in the east-west direction, which may indicate its rotation axis is north-south, coincident with the field lines at the center of the envelope. Higher angular resolution observations are needed to support this interpretation. 

In addition, we report an overall linear decrease of the polarization fraction as the Stokes\,I intensity increases up to 17\mjy (Figure \ref{f:pi_mm15}). This behaviour (so-called the \lq\lq polarization hole\rq\rq) has been already seen in many other star-forming regions and it is usually interpreted in two ways \citep[e.g.,][]{2018Soam,2019LeGouellec,2019Pattle,2019Kwon,2020LeGouellec,2020Planck12}. 
Firstly, it can indicate the detection of the polarized emission produced by the magnetic alignment via radiative torques due to interstellar photons in a surface layer of the cloud. This would mean that the grains are aligned with the magnetic fields only in the outer skin of the cloud. 
Second, it can indicate several depolarization effects. These can be caused by unresolved magnetic field structure within the beam, grain growth in denser regions (and hence less alignment due to more spherical shape of the larger grains in these regions), collisional de-alignment of grains in high-density regions, destruction of grains via RAT-D \citep{2019Hoang}, or a lower flux of photons due to higher optical depth close to young stars. In the case of MM15, we have also detected a different pattern closer to the millimeter envelope peak emission (Figures \ref{f:mm15_pol} and \ref{f:pi_mm15}). Over 17\mjy (marked approximately by the $15\sigma$ contour in Fig. \ref{f:mm15_pol}) the polarization fraction increases with Stokes\,I, reverting the commented tendency. The reasons for this behaviour are exactly the opposite to those listed before. It could be due to a more efficient grain alignment (including different alignment mechanisms at play), a more organized magnetic field in smaller scales, or a larger unattenuated flux of photons \citep{2008Whittet,2018Soam}.

\subsection{The Central Shell: Condensations, Arches and Loops}\label{s:dis_arcs}
Most of the Central Shell comprises ionized gas and warm dust emitting at 1.2\,mm wavelengths. We observe a concentric structure including an inner mostly ionized region (only 23\% of the mm emission is produced by dust), a projected dusty belt-like structure (70\% of the emission produced by dust) and an external ionized region in the shape of arches and loops (50\% emission produced by dust). The inner region contains three compact condensations (S11, S12 and S14). While S14 is primarily made of ionized gas, S11 and S12 could harbour up to $\sim1$\msun of dust and gas. These condensations could be related with material dragged by the outflows or the explosive event, but we can not reject that S11 and S12 contain the circumstellar envelope of very embedded protostellar candidates, undetected at infrared or optical wavelengths, and located (in projection) close to the center of the shell. 

Along the belt-like structure, we identified ten millimeter condensations. We estimate that $\sim75$\% of their emission stems from thermal dust. Interestingly, the position of the Feldt's star lies within this belt of dust, at about 3000\,au from the center of the shell. In the explosive scenario, the members of the original cluster have blew away from the center as happened with Orion I source and the Becklin-Neugebauer object \citep{1995Plambeck,2005Bally,2005Rodriguez,2009Zapata,2017Rodriguez,2017Luhman,2020Rodriguez}. This seems like a plausible origin of the dust belt as well, that may embed the emission from other protostars blew away in the dispersal event \citep{2006Puga}. In fact, we estimate a dust and gas mass of 55-115\msun assuming a dust temperature 75-150\,K. This may correspond with the remnant of an envelope which $\sim$1000\,years ago harboured a high-mass star-forming protocluster. Assuming the belt material is expanding with a velocity of 35\kms \citep{1998Acord}, the shell may possess a kinetic energy 2.5-5.0$\times10^{48}$\,ergs and a dynamical time of 690 years (note that the velocity of expansion has not been accurately measured yet). These values are in good agreement with others found in the literature (see Section \ref{s:intro}).

\begin{figure*}
\minipage{0.49\textwidth}
  \includegraphics[width=\linewidth]{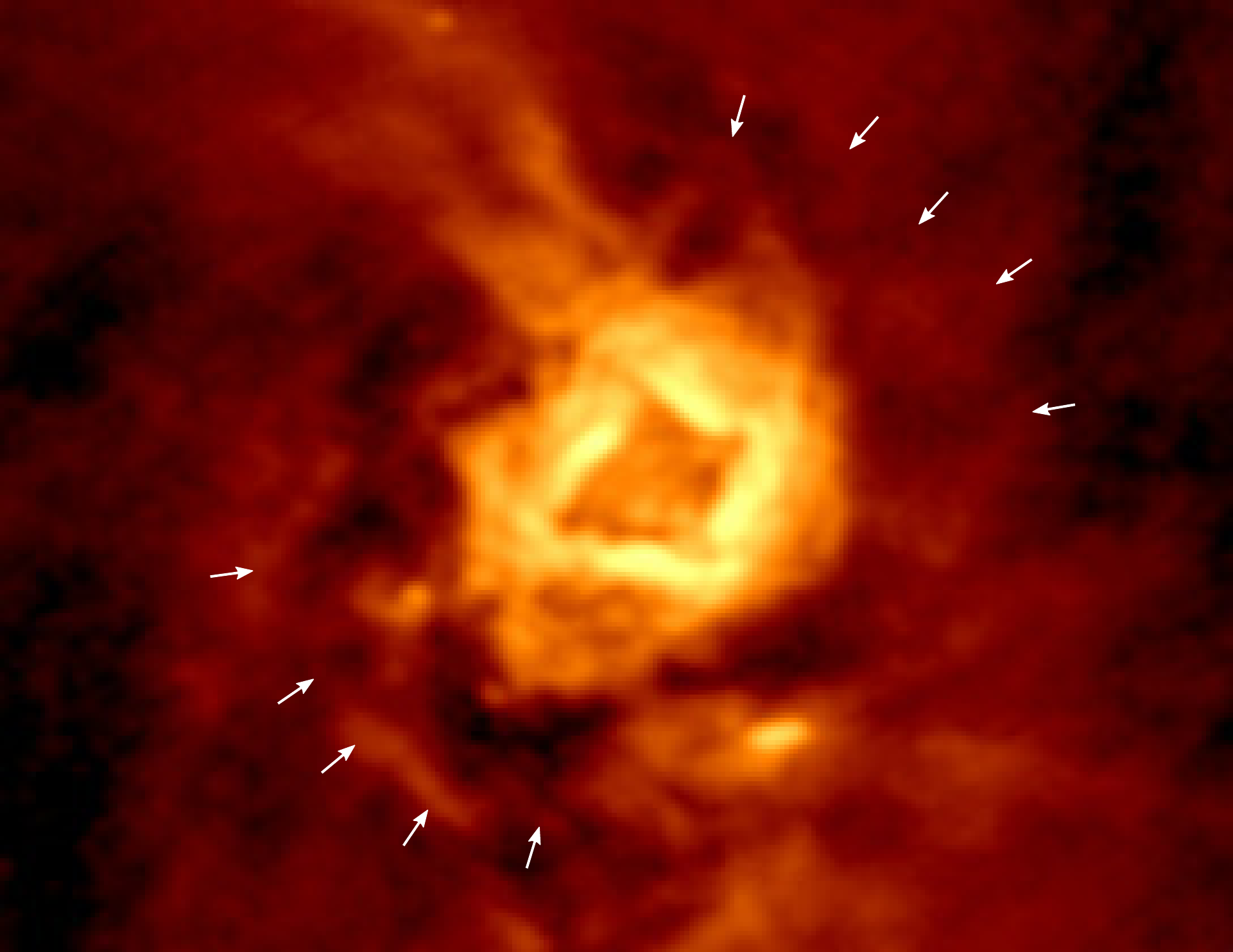}
\endminipage\hfill
\minipage{0.49\textwidth}
  \includegraphics[width=\linewidth]{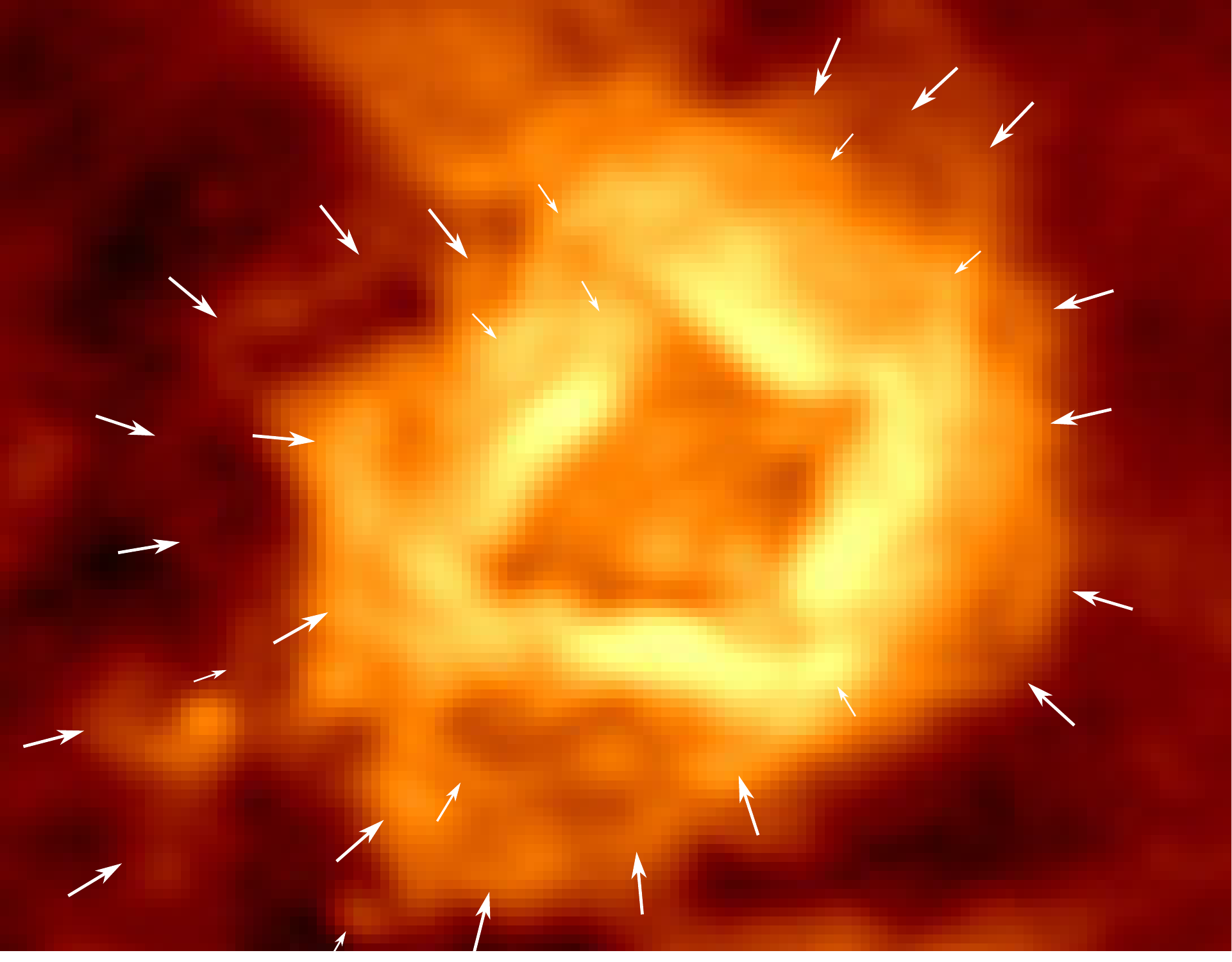}
\endminipage
  \caption{ALMA 1.2\,mm continuum emission. Identification of arches and structures in the outer (left) and inner (right) rings. We used logarithmic color stretchings to highlight the complex of clumpy arches and loops.}
  \label{f:archid}
\end{figure*}

Beyond the dusty belt there are nine more millimeter compact sources: S1, S2, S3, S7, S9, MM2, MM4, MM5 and MM7. Some of these sources could harbor protostellar envelopes, but most seem likely related with a lattice of partially ionized gas mixed with dust arches that surround the dusty belt. We have marked with arrows some of the structures of this lattice in Figure \ref{f:archid}. Many of them are in the form of arches pointing away from the central shell, which may suggest that they could be produced by gas and dust dragged-away by the ejecta from the members of the original protocluster and the explosive event. One possibility is that the arches are tracing the bow-shocks created by the outflows that may have been launched before or during the explosive event, but we have not found many matches between them and the CO outflows from \cite{2020Zapata}. The gas in these structures is partially ionized by the high energy photons from the protocluster.

In the left panel of Figure \ref{f:archid} we mark several arches or patches of arches more or less concentric, with distances between $1\farcs6$ and $4\farcs0$ from the shell center. These include previously reported northwest and southeast partially ionized loops also seen in radiowavelengths \citep[Fig. \ref{f:free-free}; see also][]{2008Hunter}. In the right panel of Figure \ref{f:archid} it is possible to identify a few more distant (at about $5\arcsec$ from the center of the shell) and fainter loops and arches. Thus, the ALMA continuum observations seem to reveal two different rings of arches or structures: an inner ring comprising a more isotropic distributed set of structures, and an outer ring showing a wide bipolar northwest/southeast distribution. We have checked for spatial coincidences between these arches and the tip of the 34 outflowing filaments produced in the explosive event \citep{2019Zapata,2020Zapata}. In general, the tips of the explosive outflows are much farther away from the center of the shell. However, while most these CO filaments do not match the arches, there is one with a clear match about 2$\arcsec$ northeast of MM4, due east from the center of the shell (Fig. \ref{f:overview}). This suggests that arches and explosive outflows are not intimately linked, but may have been produced in a similar manner. Appendix \ref{s:appendix} describes a speculative scenario interpreting some of these structures.  
%The arch lattice is not that extended, indicating that the ionizing source of the inner ring structures may be in the neighborhood of the Central Shell, probably being the O5V Feldt star.

\section{Summary and Conclusions}\label{s:conclusions}

In this work, we have presented new deep 1.2\,mm observations carried out with the ALMA telescope as part of the MagMaR survey. At spatial scales of 10,000\,au, the data revealed a large filamentary dust structure broken up in two pieces by the central UCHII region. The masses of the two portions of the large-scale filament are 125 and 145\msun each (adopting a dust temperature of 15\,K), and the mass of the Central Shell, whose millimeter emission is contributed 40\% by ionized emission, is estimated in 55-115\msun. At spatial scales of $\lesssim 1,000$\,au we found 23 dust millimeter envelopes mostly embedded within the North and South Filaments with masses ranging between 0.2 and 10.2\msun. Delineating the Central Shell, a projected belt-like structure of 4,500\,au in radius also harbours several millimeter condensations. About 75\% of their emission stems from thermal dust and we estimate their masses to be between 0.7 and 8.6\msun. In total, we found up to 18 of these partially ionized condensations within the Central Shell boundaries. The shape of the dusty belt is not circular or even regular and it is surrounded by tens of arches and loops comprising partially ionized gas. Many of these arches are shaped like bow-shocks pointing radially away from the center of the UCHII region. We speculate that they may be related with the ejecta before and after the explosive dispersal event ocurred.  
%As a proof of concept, we identified arches using very eccentric ellipses sharing the same center. There appear to be at least two rings of arches. The innermost (radii$\lesssim10,000$\,au), daisy-like, comprises an almost isotropic ensemble of arches/bows. Most of these bows have a bipolar counterpart. We found that the outer ring of arches (radii$\simeq15,000$\,au) is distributed in a wide northwest-southeast bipolar pattern.

The fraction of the millimeter polarized emission observed by ALMA is $\sim4.4$\% in the filaments and $\sim2.1$\% at the Central Shell, reaching a minimum of $\sim0.35$\% at the dusty belt position (after accounting for the free-free contamination at millimeter wavelengths). We interpret the polarization originated by dust aligned by magnetic fields in the filaments. We found that the magnetic field is aligned along and orthogonal to the North and South Filament directions, respectively. The magnetic field pattern toward the Central Shell seems to be mainly radial and mostly detected toward the western hemisphere. The radial distribution of the magnetic field could be a signpost linked with explosive dispersal events, since it has also been found in Orion BN/KL. We also compared this radial pattern with that observed in supernova remnants and recall that \target coincides with a high-energy gamma-ray source. Finally, we also reported an abrupt $\sim90\degr$ change in the field lines toward the center of the most massive millimeter envelope found in the region (MM15), accompanied by a progressive decrease of the polarization fraction with a sudden turn toward the center of the envelope.

\acknowledgments
%We are very grateful to the anonymous referee for all the comments and suggestions that helped a lot to improve the text and contents of this work. 

M.F.L. and I.S. acknowledge the fruitful discussions of a polarization group comprised by Haifeng Yang, Zhi-Yun Li, Leslie Looney, Daniel Lin and Rachel Harrison. P.S. was partially supported by a Grant-in-Aid for Scientific Research (KAKENHI Number 18H01259) of Japan Society for the Promotion of Science (JSPS). L.A.Z. acknowledges financial support from CONACyT-280775 and UNAM-PAPIIT IN110618 grants, M\'exico.  C.L.H.H. acknowledges the support of the NAOJ Fellowship and JSPS KAKENHI grants 18K13586 and 20K14527. J.M.G. is supported by the grant AYA2017-84390-C2-R (AEI/FEDER, EU). K.T. was supported by JSPS KAKENHI Grant Number 20H05645.

This paper makes use of the following ALMA data:
ADS/JAO.ALMA\#2017.1.00101.S. ALMA is a partnership of
ESO (representing its meMRr states), NSF (USA) and NINS (Japan), together
with NRC (Canada), NSC and ASIAA (Taiwan), and KASI (Republic of Korea), in
cooperation with the Republic of Chile.  The Joint ALMA Observatory is
operated by ESO, AUI/NRAO and NAOJ. The National Radio Astronomy Observatory is a facility of
the National Science Foundation operated under cooperative agreement by Associated Universities, Inc.

%It is also partially based on observations made with the NASA/ESA Hubble Space Telescope, and obtained from the Hubble Legacy Archive, which is a collaboration between the Space Telescope Science Institute (STScI/NASA), the Space Telescope European Coordinating Facility (ST-ECF/ESA) and the Canadian Astronomy Data Centre (CADC/NRC/CSA).

\facility{ALMA, HST}
\software{DS9 \citep{2003Joye}, CASA \citep[v5.6.2][]{2007McMullin}, Astropy \citep{2013Astropy}, Karma \citep{1995Gooch}}

\newpage
\bibliography{biblio}

\appendix 
\section{Speculative Scenario Explaining the Set of Arches and Loops} \label{s:appendix}
\begin{figure*}
\minipage{0.33\textwidth}
  \includegraphics[width=\linewidth]{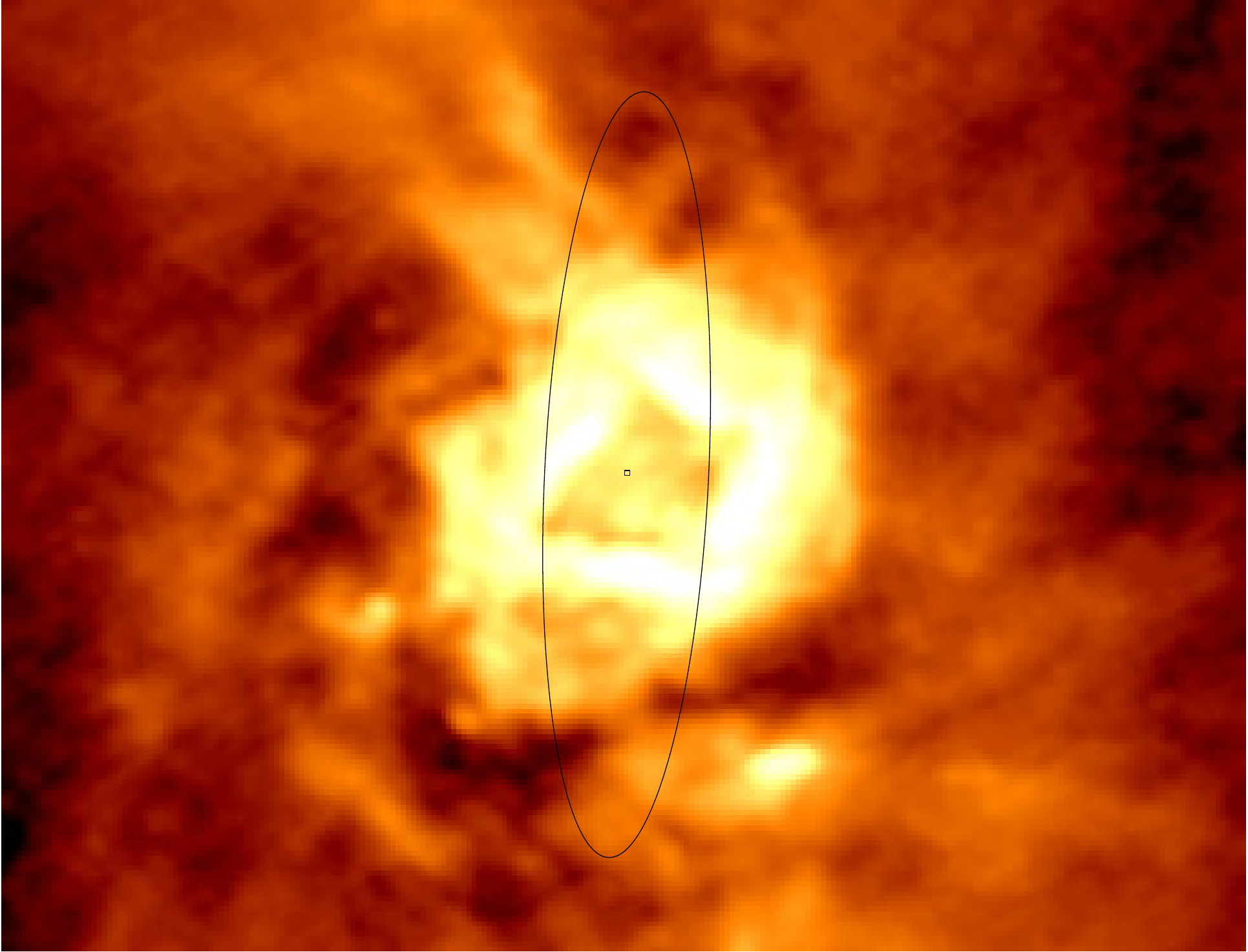}
\endminipage\hfill
\minipage{0.33\textwidth}
  \includegraphics[width=\linewidth]{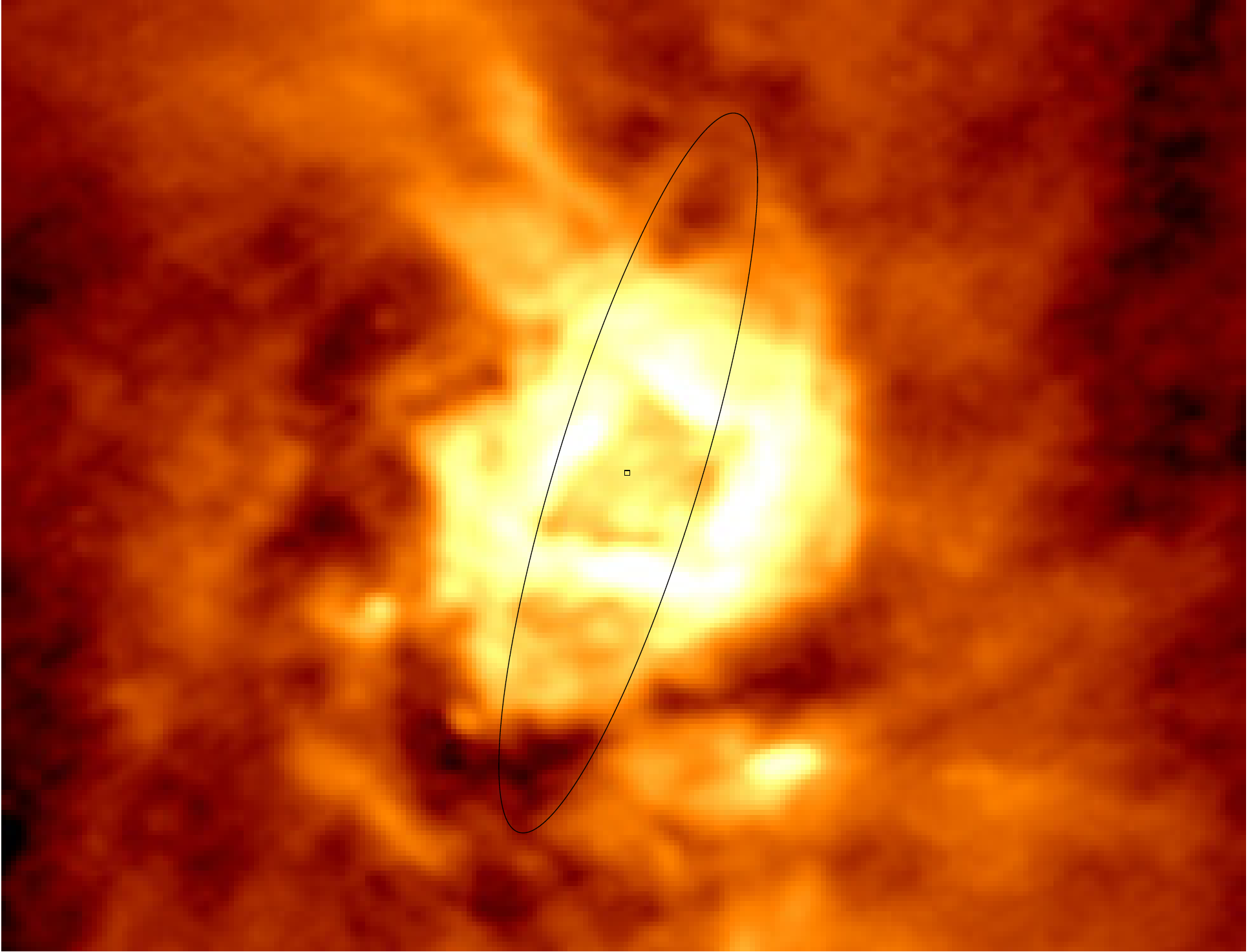}
\endminipage\hfill
\minipage{0.33\textwidth}
  \includegraphics[width=\linewidth]{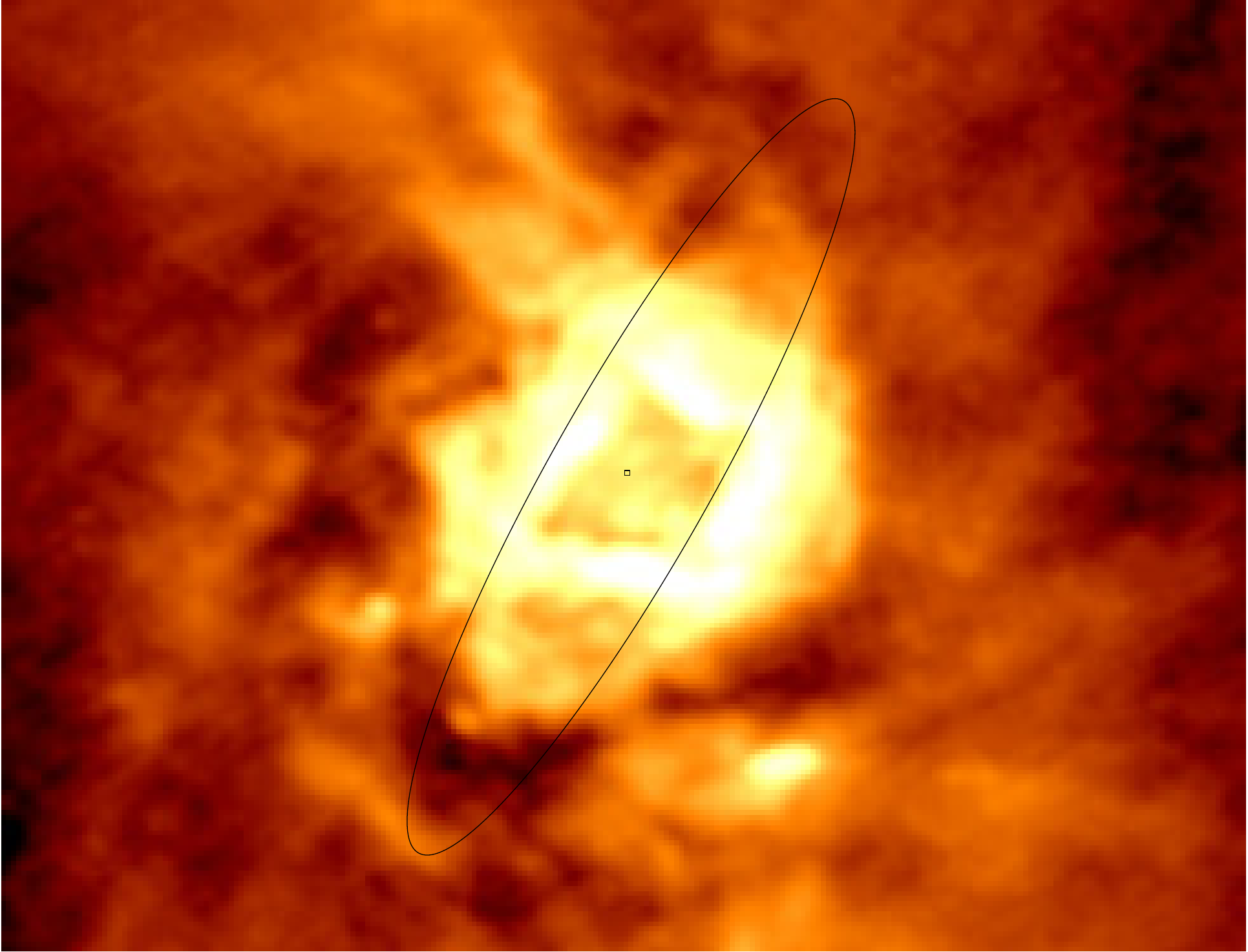}
\endminipage\hfill
\\
\minipage{0.33\textwidth}
  \includegraphics[width=\linewidth]{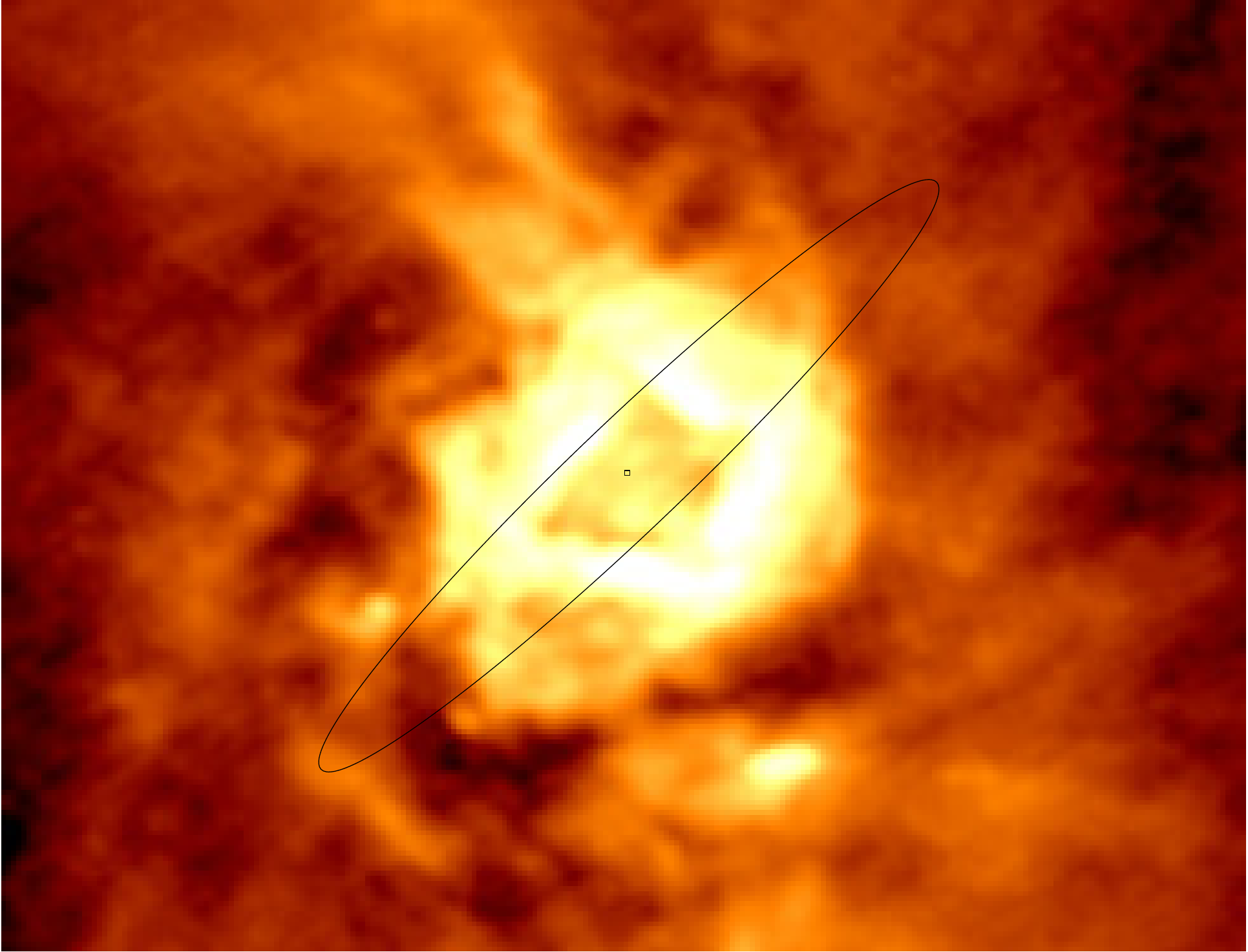}
\endminipage\hfill
\minipage{0.33\textwidth}
  \includegraphics[width=\linewidth]{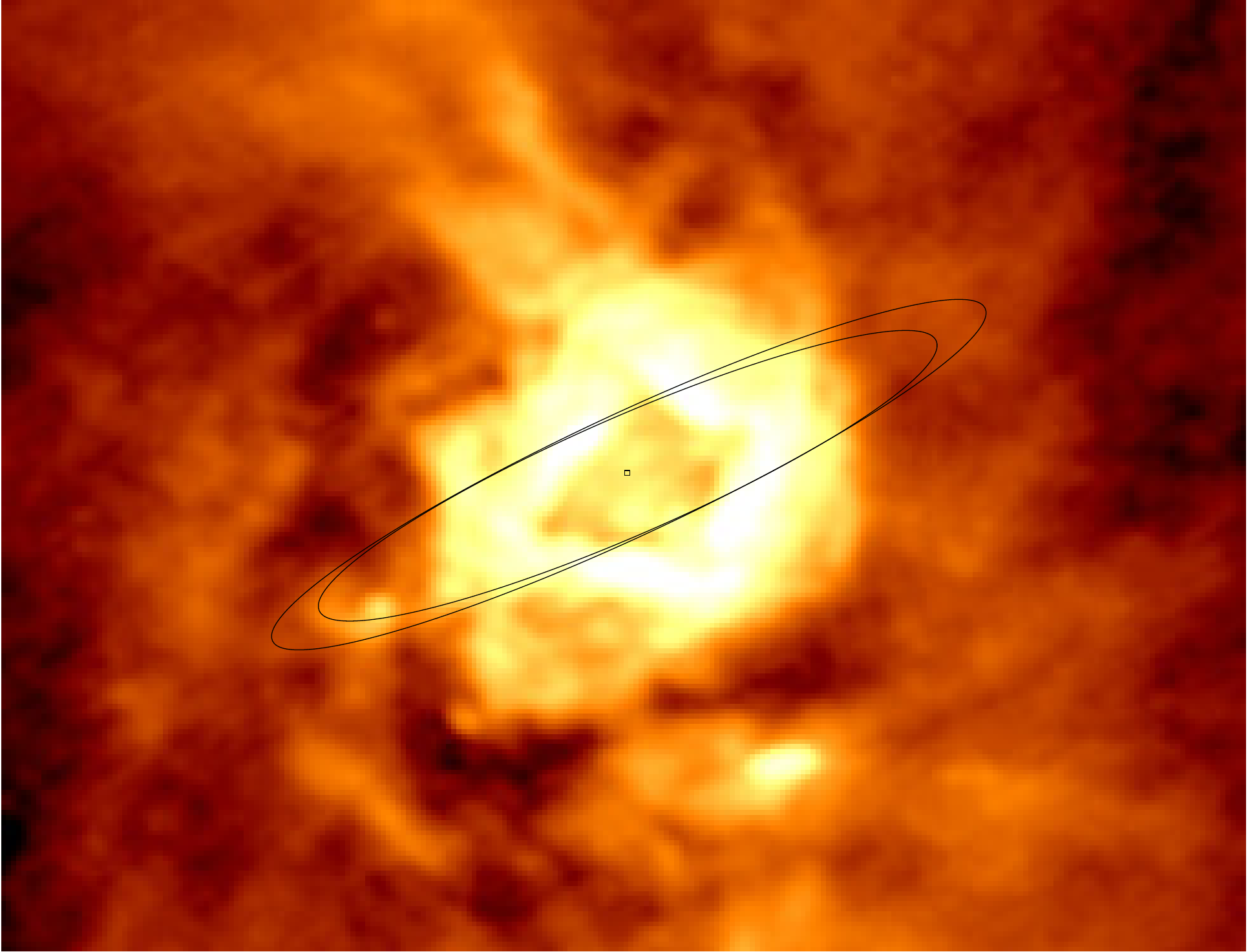}
\endminipage\hfill
\minipage{0.33\textwidth}
  \includegraphics[width=\linewidth]{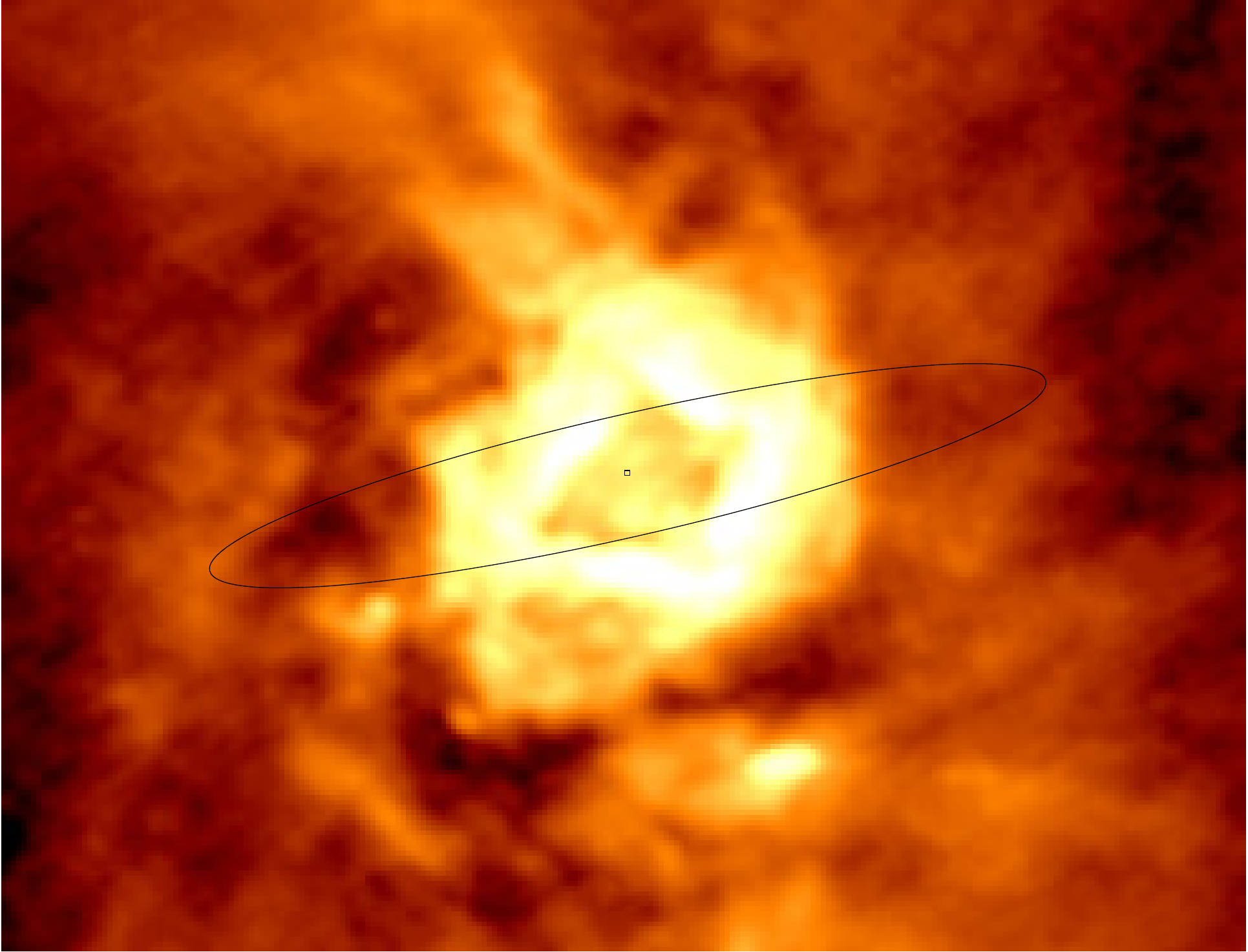}
\endminipage\hfill
  \caption{Proof of concept of ellipse identification in the outer ring of arches and structures of \target.}
  \label{f:archid2}
\end{figure*}

As a speculative interpretation of the complex of arches and loops found in the continuum emission, we include in Figure \ref{f:archid2}, as a proof of concept, a set of ellipses sharing the same center and having about the same size but different orientations overlapped with the millimeter continuum emission. The tips of some of the ellipses appear to trace the position of some of the arches, as if these were produced by bipolar structures sharing a common center (i.e., some arches appear as pairs of opposite bipolar structures). Pulsating outflows have been reported in Cepheus\,A\,HW2 \citep{2009Cunningham,2013Zapata}, S140 \citep{2002Weigelt} and IRAS\,15398-3359 \citep{2021Vazzano}. The pulsations are thought to be produced by the tidal gravitational pull that a companion in an eccentric orbit exerts on the disk/protostar system driving the outflow when approaching the periastron. With each passage, the disk plus protostar system is tilted and the accretion boosted, causing the launching of new outflows in different directions, in principle, perpendicular to the disk.  One possibility for \target is that the arches from the outer ring are the last remnants of the outflows of one such system. At a certain moment the orbital motions of the multiple system would become more chaotic and the interactions stronger, with faster and faster periastron passages producing more frequent and disorganized tilts and consequent ejections in directions that may result in the observed ensemble of bow-shock arches. These may be the last bipolar ejections before some members of the multiple system (and their disks) get close enough to unleash an isotropic blast which may impulse high velocity fragments resulting in outflowing filaments at hundreds of \kms, and the expansion of a large mass of compacted dust and gas that may explain the shell of debris seen as a belt-like structure in projection. As mentioned above, this explanation is at this moment very speculative and more evidence (e.g., gas kinematics and chemistry, identification of possible protostars, work simulating this scenario) should be gathered to sustain it. However, a similar scenario has been recently proposed to explain the Homunculus nebula and the wind ejecta surrounding Eta Carinae following a stellar merger in an unstable triple system \citep[][and references therein]{2020Hirai}.

\end{document}